\documentstyle[psfig]{mn}

\newif\ifAMStwofonts

\def\beq{\begin{equation}}
\def\eeq{\end{equation}}
\def\bea{\begin{eqnarray}}
\def\eea{\end{eqnarray}}
\def\cir{\epsilon}
\def\df{{\rm DF }}
\def\c{{\rm c}}
\def\ltsima{$\; \buildrel < \over \sim \;$}
\def\simlt{\lower.5ex\hbox{\ltsima}}
\def\gtsima{$\; \buildrel > \over \sim \;$}
\def\simgt{\lower.5ex\hbox{\gtsima}}
\def\c{c_{\rm h}}
\def\apj{ApJ}
\def\aj{AJ}

\ifoldfss

  \newcommand{\bld}[1] {{\bf #1}}
  \ifCUPmtlplainloaded \else
    \NewTextAlphabet{textbfit} {cmbxti10} {}
    \NewTextAlphabet{textbfss} {cmssbx10} {}
    \NewMathAlphabet{mathbfit} {cmbxti10} {} 
    \NewMathAlphabet{mathbfss} {cmssbx10} {} 
  \fi
  \ifAMStwofonts
    \ifCUPmtlplainloaded \else
      \NewSymbolFont{upmath} {eurm10}
      \NewSymbolFont{AMSa} {msam10}
      \NewMathSymbol{\upi}     {0}{upmath}{19}
      \NewMathSymbol{\umu}     {0}{upmath}{16}
      \NewMathSymbol{\upartial}{0}{upmath}{40}
      \NewMathSymbol{\leqslant}{3}{AMSa}{36}
      \NewMathSymbol{\geqslant}{3}{AMSa}{3E}

       \let\ge=\geqslant
    \fi
  \fi
\fi 

\ifnfssone
  \newmathalphabet{\mathit}
  \addtoversion{normal}{\mathit}{cmr}{m}{it}
  \addtoversion{bold}{\mathit}{cmr}{bx}{it}

  \newcommand{\bld}[1] {\mathbf{#1}}
  \newmathalphabet{\mathbfit} 
  \addtoversion{normal}{\mathbfit}{cmr}{bx}{it}
  \addtoversion{bold}{\mathbfit}{cmr}{bx}{it}
  \newmathalphabet{\mathbfss} 
  \addtoversion{normal}{\mathbfss}{cmss}{bx}{n}
  \addtoversion{bold}{\mathbfss}{cmss}{bx}{n}
  \ifAMStwofonts
    \ifCUPmtlplainloaded \else
      \UseAMStwoboldmath
      \makeatletter
      \new@mathgroup\upmath@group
      \define@mathgroup\mv@normal\upmath@group{eur}{m}{n}
      \define@mathgroup\mv@bold\upmath@group{eur}{b}{n}
      \edef\UPM{\hexnumber\upmath@group}
      \new@mathgroup\amsa@group
      \define@mathgroup\mv@normal\amsa@group{msa}{m}{n}
      \define@mathgroup\mv@bold\amsa@group{msa}{m}{n}
      \edef\AMSa{\hexnumber\amsa@group}
      \makeatother
      \mathchardef\upi="0\UPM19
      \mathchardef\umu="0\UPM16
      \mathchardef\upartial="0\UPM40
      \mathchardef\leqslant="3\AMSa36
      \mathchardef\geqslant="3\AMSa3E

       \let\ge=\geqslant
    \fi
  \fi
\fi 

\ifnfsstwo

  \newcommand{\bld}[1] {\mathbf{#1}}
  \DeclareMathAlphabet{\mathbfit}{OT1}{cmr}{bx}{it}
  \SetMathAlphabet\mathbfit{bold}{OT1}{cmr}{bx}{it}
  \DeclareMathAlphabet{\mathbfss}{OT1}{cmss}{bx}{n}
  \SetMathAlphabet\mathbfss{bold}{OT1}{cmss}{bx}{n}
  \ifAMStwofonts
    \ifCUPmtlplainloaded \else
      \DeclareSymbolFont{UPM}{U}{eur}{m}{n}
      \SetSymbolFont{UPM}{bold}{U}{eur}{b}{n}
      \DeclareSymbolFont{AMSa}{U}{msa}{m}{n}
      \DeclareMathSymbol{\upi}{0}{UPM}{"19}
      \DeclareMathSymbol{\umu}{0}{UPM}{"16}
      \DeclareMathSymbol{\upartial}{0}{UPM}{"40}
      \DeclareMathSymbol{\leqslant}{3}{AMSa}{"36}
      \DeclareMathSymbol{\geqslant}{3}{AMSa}{"3E}

       \let\ge=\geqslant
    \fi
  \fi
\fi 

\ifCUPmtlplainloaded \else
  \ifAMStwofonts \else 
    \def\upi{\pi}
    \def\umu{\mu}
    \def\upartial{\partial}
  \fi
\fi
\title[On Life and Death]
{On the Life and Death of Satellite Haloes}
 
\author[G. Taffoni et al.]{Giuliano Taffoni $^1$, Lucio Mayer $^2$, 
Monica Colpi $^3$, Fabio Governato $^{2,4}$ \\
$^1$ SISSA, via Beirut 4 - 34014 Trieste, Italy, taffoni@sissa.it\\
$^2$ Department of Astronomy, University of Washington,
Seattle, WA, 98195 USA, mayer@astro.washington.edu\\
$^3$Dipartimento di Fisica, Universit\`a Degli Studi di
Milano Bicocca, Piazza della Scienza 3, I-20126 Milano, Italy, 
colpi@mib.infn.it\\
$^4$ Osservatorio Astronomico di Brera,
via Brera 28, 20121 Milano - Italy, fabio@astro.washington.edu}

\date{Submitted to MNRAS, September 2002}

\pagerange{\pageref{firstpage}--\pageref{lastpage}}
\pubyear{1994}
\begin{document}
\maketitle
\label{firstpage}
\begin{abstract}

We study the evolution of dark matter satellites orbiting 
inside more massive haloes using semi-analytical tools coupled 
with high-resolution N-Body simulations. We select 
initial satellite sizes, masses, orbital energies, and eccentricities as
predicted by hierarchical models of structure formation.
Both the satellite (of initial mass $M_{\rm s,0}$) and the main halo 
(of mass $M_{\rm h}$) are described by a Navarro, Frenk \&
White density profile with various concentrations.

We explore the interplay between dynamical friction and tidal mass
loss/evaporation in determining the final fate of the satellite.  We
provide a user-friendly expression for the dynamical friction
timescale $\tau _{\rm df,live}$ and for the disruption time for a
{\it live} (i.e. mass losing) satellite.This can be easily implemented into existing semi-analitycal models of galaxy formation improving considerably the way they describe the evolution of satellites.

Massive satellites ($M_{\rm s,0}>0.1 M_{\rm h}$) starting from typical 
cosmological orbits sink rapidly (irrespective of the
initial circularity) toward the center of the main halo where they
merge after a time $\tau_{\rm df,rig}$, as if they were
rigid.  Satellites of intermediate mass ($0.01 M_{\rm h}<M_{\rm
s,0}<0.1 M_{\rm h}$) suffer severe tidal mass losses as dynamical
friction reduces their pericenter distance. In this
case mass loss increases substantially their decay time with respect to
a rigid satellite. The final
fate depends on the concentration of the satellite, $c_{\rm s}$, 
relative to that of the main halo,  $c_{\rm h}$.
Only in the unlikely case where $c_{\rm s}/\c \simlt 1$  satellites are 
disrupted. In this mass range, $\tau_{\rm df,\rm live}$ gives a measure 
of the merging time. Among the satellites whose orbits decay
significantly, those that survive must have been moving preferentially on more
circular orbits since the beginning as dynamical friction does not induce
circularization.
Lighter satellites ($M_{\rm s,0}< 0.01 M_{\rm h}$) do not suffer
significant orbital decay and tidal mass loss stabilizes even further
the orbit.  Their orbits should map those at the time of entrance into the main halo.

After more than a Hubble time satellites have masses $M_{\rm s}\sim
1-10\%M_{\rm s,0},$ typically, implying $M_{\rm s}< 0.001 M_{\rm h}$
for the remnants. In a Milky Way like halo, light 
satellites should be present even after several orbital times
with their baryonic components
experimenting morphological changes due to tidal stirring.

They coexist with the remnants of more massive satellites depleted
in their dark matter content by the tidal field, which should move
preferentially on tightly bound orbits.

\end{abstract}


\section{Introduction.}
In the current view, structure formation in the Universe proceeds
through a complex hierarchy of mergers between dark matter haloes, from
the scale of dwarf galaxies up to that of galaxy clusters. Galaxy
formation occurs within dark matter haloes while these 
evolve and grow through a series of mergers. 
During the assembly of these systems, various processes, like
morphological transformations of the stellar and gaseous components are
expected to happen. Therefore, understanding the dynamical evolution
of dark matter haloes is a fundamental step of any theory of galaxy
formation.

N-body simulation are widely used to study the dynamical evolution of
cosmic structures (Governato et al.  1999)
and they have been the most useful tool to address this
problem, so far. Studying in detail the internal
dynamical evolution of many haloes requires however a high number of particles
in order to resolve substructure avoiding its artificial
evaporation (Moore, Katz \& Lake 1996; Ghigna et al, 1998; Moore et
al. 1999; Lewis et al. 2000; Jing \& Suto 2000; Fukushige \& Makino
2001). The heavy computational burden associated with such
cosmological runs limits the level of detail at which the evolution
of the internal structure of satellites can be followed.  On the other
end, non-cosmological simulations at very high resolution, 
for a limited number of
systems, have shown that merging and other
interactions between haloes (and eventually between their embedded luminous
galaxies) like harassment and tidal stirring can dramatically affect
their global properties and their internal structure 
(Huang \& Carlberg 1997; Naab, Burkert \&
Hernquist 1999; Vel\'azquez \& White 1999; Moore et al. 1996, 1998;
Mayer et al. 2001a,b, Zhang et al. 2002).

A different approach to the problem of structure formation and
evolution is brought about by semi-analytical methods. The backbone of
the semi-analytical models of galaxy formation (Somerville \& Primack
1999; Kauffmann et al. 1999; Cole et al. 2000) is the merging history
of dark matter haloes which can be Monte-Carlo generated (Somerville
\& Kolatt 1999; Sheth \& Lemson 1999; Cole et al. 2000) or calculated
from N-Body simulations (Kauffmann, White
\& Guiderdoni, 1993). The evolution of  substructures in
semi-analytical models is followed in a simplified way: a
merging event between unequal mass haloes takes place when the lighter
halo reaches the center of the more massive one. The time scale for
this to happen is obtained from the local application of
Chandrasekhar's formula (1943) for dynamical friction.

However, as the magnitude of the frictional drag depends on the mass
of the satellite and this is a time-dependent quantity, we expect
stripping to ultimately affect the orbital decay rate.  Somerville \&
Primack (1999) include a simple recipe which accounts for mass
stripping, reducing the mass of the satellite by re-calculating its
tidal radius while it spirals toward the center along a circular
orbit.

Colpi, Mayer \& Governato (1999, hereafter CMG99) quantified the
interplay between dynamical friction and tidal stripping for a
selected sample of orbits and satellites mass.  Using high-resolution
N-Body simulations, they showed that small satellites (with initial
masses 50 times smaller than that of the primary) undergo tidal mass
loss and their orbit decay as if they had an ``effective mass'' $\sim
60$ per cent lower than the initial; on typical cosmological orbits
they never merge at the center of the primary because the magnitude of
the drag is drastically reduced at such a small effective mass.  The
fraction of mass lost by the satellite is strictly related to the
particular orbital parameters and halo profile assigned to the haloes.
In order to improve this recipe and make it more physically motivated
it is necessary to recognize that mass loss is the consequence not
only of the initial tidal truncation but also of the repeated
gravitational shocks occurring at each pericenter passage (Taylor \&
Babul 2000, TB; Gnedin, Hernquist \& Ostriker 1999, GHO; Weinberg
1994); the strength and effectiveness of the shocks depends on the
central density profile and orbit of the satellite and might lead to
its complete disruption before the merger is completed.  This regime
of disruption or, at least, of mass evaporation during orbital decay,
is completely neglected by semi-analytical models of galaxy formation,
even by the recipe adopted by Somerville \& Primack (1999). However it
has been shown that satellite orbits are very eccentric in CDM models
(e.g. Tormen 1997; Ghigna et al. 1998), and this points to strong
tidal shocks.

The full dynamical evolution of the satellites must be studied using
haloes similar to those forming in cosmological simulations, that have
cuspy density profiles.  In this work we will consider haloes with
Navarro, Frenk \& White (1996; 1997; hereafter NFW) profiles as
opposed to the previous work, where our analysis was restricted to
isothermal spheres with cores (CMG99). We note that more recent higher
resolution simulations (Moore et al. 1999; Ghigna et al. 2000; Bullock
et al. 2000; Jing \& Suto 1999; Governato, Ghigna \& Moore 2001) find
that the inner slope of the density profile is even steeper than the
NFW.

Following the same philosophy of CMG99, we use semi-analytical methods
to describe the orbital evolution and mass loss of satellites in an
NFW profile, and we compare the results with high resolution N-Body
simulations. In particular, we will use the theory of linear response
to model dynamical friction and study orbital decay (Colpi 1998; Colpi
\& Pallavicini 1998).  We will apply the
theory of gravitational shocks developed by GHO to model tidal mass
loss and the disruption of satellites (Taylor \& Babul 2001; Hayashi
et al. 2002).

The paper is organized as follow: we first review the main features of
NFW haloes (Section 2), and of the drag force as derived using the
theory of linear response (Section 2.1). In Section 3 we study the
orbital decay of a rigid satellite. We then move on describing the
effects of the tidal perturbation both when the orbit is stable and
when it decays due to dynamical friction (Section 4). Finally we
discuss the global effect of dynamical friction (DF) and mass loss on
the evolution of satellites.

\section{Orbital Evolution in a NFW Profile.}
A realistic representation of the density profile consistent with the
findings of structure formation simulations is needed for a meaningful
study of the disruption of satellites.
Here, we  use the so called 'universal density profile' of Navarro,
Frenk \& White (1996):
\beq
\label{eq:rho}
\rho(r)= {M_{\rm h} \over 4 \pi R_{\rm h}^3}  
{\delta_c \over (c_{\rm h} \, x) (1+c_{\rm h}\,x)^2} \;,
\eeq 
where $x=r/R_{\rm h}$ is the dimensionless radius in units of the 
virial radius $R_{\rm h}$, $M_{\rm h}$ is the mass of the halo inside
$R_{\rm h},$ $c_{\rm h}=r_{\rm s}/R_{\rm h}$ 
is the concentration parameter ($r_{\rm s}$ is a
scale radius), and $\delta_{\rm c}=
c_{\rm h}^3/[\ln(1+c_{\rm h})-c_{\rm h}/(1+c_{\rm h})].$

A halo of given mass and size does not have a unique NFW profile; the
concentration $c$ plays the role of a free parameter that basically
tells how much of the total mass is contained within a given inner
radius.  Haloes with a higher concentration have more mass in the
central part and should thus be more robust against tidal effects.  We will
consider various concentrations for both the primary halo and the
satellite.

The mass profile of a spherically symmetric halo (i.e. the mass
contained inside a sphere of radius $r$) can be obtained integrating
equation~(\ref{eq:rho}) over the spherical volume
\beq 
M(r)=M_{\rm h} {\ln(1+c_{\rm h}\,x)-c_{\rm h}\,x /( 1+c_{\rm h}\,x) \over
\ln(1+c_{\rm h})-c_{\rm h}/(1+c_{\rm h})} \;,  
\label{eq:nfwmass}
\eeq 
and used to calculate the circular velocity profile, $V_{\rm
c}^2(r)= G M(r)/ r,$  and  the one-dimensional velocity dispersion
$\sigma(r)$  
\begin{eqnarray}
\sigma^2(r)=75.53 \; V^2_{\rm c}(2.15 R_{\rm h}/c_{\rm h}) \;(c_{\rm h}x)(1+c_{\rm h}x)^2
 \; {\cal I}(c_{\rm h} x) \\
\nonumber 
{\cal I}(x)=\int _x ^\infty \left[ \frac {\ln(1+y)}{y^3(1+y)^2}-{1
\over y^2(1+y)^3}\right]\;dy\;,
\end{eqnarray}
(Kolatt et al. 2000). 

The gravitational potential of a NFW halo can be written as:
\beq 
\phi(r)= -V^2_{\rm c}(r)+ V_{\rm h}^2
\; { c_{\rm h}/(1+c_{\rm h}) - c_{\rm h} / (1+c_{\rm h}x) \over \ln(1+c_{\rm h})-c_{\rm h}
/(1+c_{\rm h})}\;,  
\eeq 
here $V_{\rm h}$ is the value of the circular velocity at the virial
radius $R_{\rm h}$.  The orbits in this potential can be determined
using the planar polar coordinates $r(t)$ and $\theta(t),$ solving for
the equation of motion (Binney \& Tremaine 1987).  The motion of a
satellite is then determined by the initial specific angular momentum
$J$ and orbital energy $E,$ or equivalently by the radius $r_{\rm
c}(E)$ of the circular orbit having the same energy $E,$ and by the
circularity $\epsilon=J/J_{\rm c}$, where $J_{\rm c}=V_{\rm h}(r_{\rm
c}) \cdot r_{\rm c}(E)$.

We define a generalized orbital eccentricity:
\beq
\label{eq:ecc}
e={r_{\rm apo}-r_{\rm per} \over r_{\rm apo}+r_{\rm per}}\;,
\eeq 
here $r_{\rm apo}$ and $r_{\rm per}$ are the roots of the orbit equation:
\beq
\label{eq:orbiteq}
{1 \over r^2} + {2[\phi(r)-E] \over J^2}=0
\eeq
(Binney \& Tremaine 1987), and they are respectively the apocenter and
the pericenter radii of the orbit.  Using the previous equation it is
possible to derive a relation between $e$ and the
orbital parameters 
\footnote {For an Singular Isothermal Profile (SIP)
the eccentricity is only a function of $\cir$ (van den Bosch et
al. 1999)},
so that for each value of $r_{\rm c}(E)$ and $\cir$ we can determine
the apocenter and pericenter distances of the orbit. We
introduce the dimensionless radius of the circular orbit $x_{\rm c}(E)
\equiv r_{\rm c}(E)/R_{\rm h}$.

From equation~(\ref{eq:orbiteq}), the orbital period is

\beq
P_{\rm orb}=2 \, \int_{r_{\rm per}}^{r_{\rm apo}} {dx \over
\sqrt{ 2 [0.5 V_{\rm c}^2(r_{\rm c})+\phi(r_{\rm c})-\phi(x)]
-J^2/x^2}}\;.
\eeq
A satellite, described by a NFW profile, has subscript $\rm s$
in  all the  corresponding halo properties. 
The satellite, before mass loss, has a mass $M_{\rm s,0}$
inside its virial radius $R_{\rm s,0}.$

\subsection{The Theory of Linear Response}
The theory of linear response (hereafter TLR) is a relatively novel
approach to the study of dynamical friction in the non-uniform stellar
background of a spherical self-gravitating halo (Colpi \& Pallavicini
1998; Colpi 1998; CMG99; see also Weinberg 1989 for a study
of dynamical friction in a self-gravitating medium). 
The dissipative force on the satellite is
computed tracing, in a self-consistent way, the collective, 
global response of
the particles to the gravitational perturbation excited by the
satellite. The force includes the tidal deformation in the density
field (absent in an infinite uniform medium), the trailing density
wave which is evolving in time, and the shift of the barycenter of the
primary. We omit here the complex expression of the force
referring to Colpi and Pallavicini (1998) and CMG99 for details.  We
only remark that the frictional force at the current satellite
position $\vec {R}(t)$ can be written formally as an 
integral upon space and time 
\beq
\vec{F}_{\df}(t) = \\
-GM_{\rm s,0}\int d_3 \vec {r}
\int_{-\infty}^t dt' \Delta \rho(\vec {r}, t-t')
{\vec{R}(t)-\vec{r}(t)\over \vert \vec{R}(t)-\vec{r}(t)\vert^3}
\label{eq:df}
\eeq
where $\Delta \rho(r,t-t')$ maps of the time dependent
disturbances in the density field created, over time, 
by the satellite in its
motion.
%
\begin{figure}
\centerline{\psfig{figure=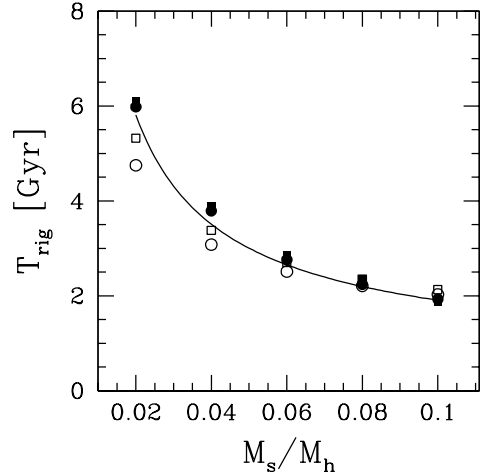,width=7.0cm}}
\caption[]{
Dynamical friction timescale $\tau_{\rm df,rig}$ versus $M_{\rm
s,0}/M_{\rm h}$ for a satellite in a Milky Way like halo with $x_{\rm
c}(E)=0.5$ and $\epsilon=1;$ filled symbols are from TLR, while open
symbols are from the local approximation of dynamical friction as
given by solving equation~(\ref{eq:nfworbeq}). Dots refer to $c_{\rm
h}=7$ while squares for $c_{\rm h}=14$; the solid line corresponds to
the fit $\tau_{\rm df,rig}\sim 1.3 R^2_{\rm h} V_{\rm h} x_{\rm c}^2/
[GM_{\rm s,0}\ln \Lambda]$ where $\ln\Lambda=\ln(1+M_{\rm h}/M_{\rm
s,0})$.  }
\label{fig:tau.vs.mass}
\end{figure}
%
\begin{figure}
\centerline{\psfig{figure=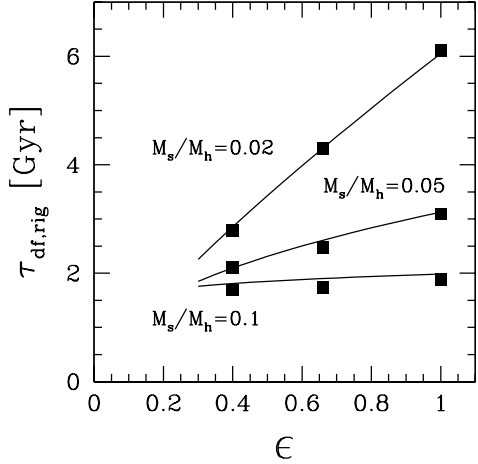,width=7.0cm}}
\caption[]{Dynamical friction time $\tau_{\rm df,rig}$ versus
circularity $\cir$ for for $M_{\rm s}/M_{\rm h}=0.02,0.05$ and 0.1,
here $x_{\rm c}(E)=0.5$.  Points are the TLR data and the solid lines
are the model results $\tau_{\rm df,rig} \propto \epsilon^\alpha$
where $\tau_{\rm df,rig}$ is given by equation~(\ref{eq:chandra}) and
we estimate $\alpha$ using equation~(\ref{eq:alfa}).  }
\label{fig:scaling.e.eps}
\end{figure}
We estimate the
force specifically for the NFW density profile.  TLR do not contain any
free parameter (no Coulomb logarithm) except the mass and the radius
of the satellite.  equation~(\ref{eq:df}) describes the sinking of
satellites moving along orbits of arbitrary eccentricity even outside
the primary halo.

\section{Numerical Simulations}
The simulations, whose results are
analyzed in the forthcoming sections of this paper, have been performed with PKDGRAV, a fast parallel binary treecode widely
used to study structure formation and galactic dynamics 
(e.g. Power et al. 2002; Ghigna et al. 1998; Mayer et al. 2001a,b; Dikaiakos
\& Stadel 1996; Stadel, 2002).
The force calculation is performed using a binary tree and using the Barnes-Hut
criterion for evaluating the multipoles up to the hexadecapole order; an 
opening angle $\theta=0.7$ was used in all the runs. The code uses a leapfrog
integrator and has multistepping capabilities. The simulations employing
a rigid satelllite use an NFW primary halo resolved by either 100.000 or 1 million particles; the satellite is modeled as in Van den Bosch et al. (1998) and
CMG99, i.e
it is represented by a point mass softened using a spline kernel (the same
kernel adopted for all the particles in the simulations).  The softening
of the particles in the primary halo is 200 pc; that of the rigid
satellite is 3.5 kpc in the reference case where the latter has a mass
$M_{\rm {s,0}}=0.05M_{\rm h}.$ 
Satellites of different masses have a
softening scaled $\sim M_{\rm {s,0}}^{1/3}$.
In the simulations for ``live'' (i.e., mass losing) satellites, 
these are resolved by either 20.000 or 50.000 particles (the same resolutions
holds in  runs of live satellites moving in a fixed external potential)
while the primary halo is resolved using 100.000 particles. 
The particle softening for satellites of different masses is 
rescaled as for the rigid satellites and the same scaling 
holds also between these particles and those of the primary halo 
(as a reference, for a satellite of 0.05 $M_{\rm h}$ the softening 
is 74.4 pc).
We note that the softening of the rigid satellite is fixed in such a way 
that a deformable satellite having high concentration ($c=20$) has roughly 
the same half mass radius than the corresponding rigid satellite of the
same mass; this ensures that a comparison between runs with rigid and
deformable satellites is meaningful (the decay rate depends on the
softening of the rigid satellite at a given mass; see e.g. Van Albada 1987).
By comparing runs with different resolutions for the primary halo
we verified the robustness of the obtained orbital decay rate 
in absence stripping.  
By comparing runs having deformable satellites with different 
resolutions we tested whether artificial heating due to two-body 
collisions was playing a role in the determining the actual mass loss
rate. 
The simulations employed timesteps as small as $10^5$ 
years in the inner regions of the halos, 
namely more than an order of magnitude smaller 
than the local orbital time; as a result, 
energy was conserved to better than 1 per cent.

\section{The Sinking of a Rigid Satellite in a NFW Profile} 

In this Section we explore the evolution of a rigid satellite of mass
$M_{\rm s,0}$ orbiting inside a halo with NFW density profile, using TLR.  The halo
is scaled to the Milky Way mass $M_{\rm h}=10^{12}M_\odot,$ has a
tidal radius $R_{\rm h}=200 \; \mathrm kpc$ and concentration $\c=7$ or
14, within the spread of cosmological values (Eke, Navarro \&
Steinmetz 2001).

Fig.~(\ref{fig:tau.vs.mass}) shows the dynamical friction time
$\tau_{\rm df,rig}$ as a function of the satellite mass $M_{\rm s,0}$
(expressed in units of $M_{\rm h}$); the satellite moves on a circular
orbit
at $x_{\rm c}(E)=0.5$ We find no
significant dependence of $\tau_{\rm df,rig}$ on the halo concentration.
The fit in the figure tries to single out the dependences 
of $\tau_{\rm df,rig}$ on
the mass of the satellite and its initial orbit in a simple way and
ties to the familiar expression of the dynamical friction timescale,
derived in the local approximation, for the case of an
isothermal sphere (Binney \& Tremaine 1987). 
If one again uses the expression of the frictional force, 
given by Chandrasekhar (1943),
treating the background 
density and dispersion velocity as local
quantities (evaluated at the satellite current position) 
\footnote{Dynamical friction 
is the result of a ``long range'' disturbance (Hernquist \& Weinberg 1989; Colpi
\& Pallavicini 1992; CMG99). Treating it as local is conceptually
incorrect. The error that results, which is
difficult to quantify unless the whole treatment is included, 
is customarily absorbed in the
Coulomb logarithm.},
the evolution equation of a satellite spiraling down
on circular orbits in a NFW main halo
is
\bea
\nonumber
\frac 1 r {d \,[r\, V_{\rm c}(r)] \over d t}=-4 \pi \ln\Lambda G^2 
M_{\rm s,0} \: {\rho(r,\c) \over V_{\rm c}^2(r)} \\
\times \left[ \rm {erf}(Y)-2 \frac Y {\sqrt {\pi}}e^{-Y^2} \right]
\label{eq:nfworbeq}
\eea  
where $Y=V_{\rm c}(r)/\sqrt 2 \sigma(r).$
This equation  can be integrated grouping all quantities depending
on $r,$ on the left hand side of equation~(\ref{eq:nfworbeq}) to give
\beq
\int_{x_{\rm c}}^0 \Theta(x,\c) \; d\,x 
= -{G M_{\rm s,0} \ln\Lambda  \over R_{\rm h}^2 V_{\rm h}} \; \tau_{\rm df,rig}\;,
\eeq
where $x_{\rm c}$ the initial radius of the circular orbit.

The function  $\Theta(x,\c)$ has an analytical expression  
that can be fitted, with an average error of one part over 1000,
as 
\beq
\Theta(x,\c) \simeq f(\c)\,x^{0.97}, 
\eeq
leading  to a dynamical friction timescale  for circular orbits  
\beq
\tau_{\rm df,rig}\sim 0.6 f(\c)\,{R^2_{\rm h} V_{\rm h}\over 
GM_{\rm s,0}}\; {x_{\rm c}^{1.97} \over \ln\Lambda }\qquad {\rm
for}\,\,\,
\cir=1 \;,
\label{eq:chandra}
\eeq
where $\ln\Lambda=\ln(1+M_{\rm h}/M_{\rm s,0})$ and $f(\c)$ is
\beq
\label{eq:fdic}
f(\c)=1.6765+0.0446 \,\c \;.
\eeq

This simple analysis explains while a fit 
similar to that for a singular isothermal sphere (see
Fig.~(\ref{fig:tau.vs.mass}) and its caption)  is acceptable even in a NFW profile.

\begin{figure}
\centerline{\psfig{figure=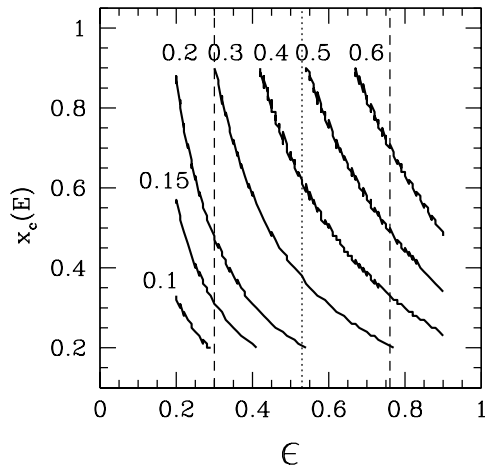,width=7.0cm}}
\caption[]{The residual mass of a satellite at the first pericenter passage as
function of the orbital parameters, when $c_{\rm s}/\c=2$. We assume
that the ``tidal cut'' instantaneously reduces the satellite mass:
Each curve is labeled with the residual mass in units of the initial
one. The vertical dotted line is the most probable value of the
eccentricity in a cosmological environment (Tormen 1997), the dashed
vertical lines are the 1$\sigma$ variance.}
\label{fig:mreps}
\end{figure}

The frictional time is also a function of the
initial orbital  circularity $\cir$:  
We thus explored the dependence of $\tau_{\rm df,rig}(\cir)$
as accretion of satellites  in a main halo
occurs preferentially
along rather eccentric orbits.  This dependence has been already
discussed in Lacey \& Cole (1993), van den Bosch et al. (1999), and
CMG99 for isothermal profiles, giving 
\beq
\tau_{\rm df,rig}(\cir) \sim 
\tau_{\rm df,rig}(\cir=1)\cir^\alpha.
\label{eq:taualpha}
\eeq  
CMG99 noticed that, for a fixed 
satellite mass ($M_{\rm s,0}\ll M_{\rm h}$), 
the timescale varies with the orbital energy, suggesting for $\alpha$
a dependence on $x_{\rm c}(E)$ (see CMG99 for the suggested values
of $\alpha$).

In a NFW halo, we find that $\alpha$ depends on $x_{\rm c}(E)$, and
on $M_{\rm s,0}/M_{\rm h}.$
We find that, whereby relatively heavy satellites decay on
a time almost independent of $\cir,$ lighter satellites decay
on much shorter times when $\cir\to 0.$ This is shown in
Fig.~(\ref{fig:scaling.e.eps}).  A useful fit to $\alpha$ as a
function of orbital energy and mass ratio is
\beq
\begin{array}{l}
\nonumber
\alpha(x_{\rm c}, M_{\rm s,0}/M_{\rm h}) \simeq  \\
0.475 \left\{ 1-\tanh \left[
10.3 \left ({M_{\rm s,0} \over M_{\rm h}}\right)^{0.33}
-7.5 x_{\rm c}\right] \right\} . 
\label{eq:alfa} 
\end{array}
\eeq
As in CMG99, we used high resolution N-body simulations to follow the
orbital decay of a rigid satellite (using up to $10^6$ particles to
sample the halo mass distribution) and found a very close match with
the theory of linear response.  The two approaches agree in a number
of details on the evolution, the most remarkable being the temporary
rise of the orbital  angular momentum observed during the final stages of the
decay. This is a manifestation of the fact that in the background
medium, no longer uniform, the satellite moves inside or close to its
distorted wake that, near pericenter distance, induces a positive
torque (Colpi et al. in preparation).

As a final remark we notice that initial eccentric orbits that decay by  dynamical 
friction do not chance significantly their degree of circularity with time, as 
was also found in isothermal profiles (van den Bosh et al. 1999; CMG99). 

\section{The Dynamical Evolution of a live Satellite}

The evolution of a rigid object is determined by the frictional 
drag force and its survival time corresponds to the dynamical friction
time $\tau_{\rm df, rig}$.  However a real satellite is not a 
rigid point mass but a deformable distribution of particles 
moving inside a halo.  Its life is then dramatically influenced by the tidal
perturbations induced by the gravitational field of the primary halo.
The global effect of the tidal perturbation is the progressive
evaporation of the satellite. This process takes place during the
orbital evolution and it is generally sensitive to the internal
properties of the satellite and of the surrounding halo.

Our aim is to model realistically the tidal effects in order to
evaluate the mass that remains bound to the satellite, $M_{\rm s}(t)$,
each time along the orbit.

We distinguish two tidal effects: a tidal truncation ({\em tidal
cut}), originated by the average tidal force exerted by the main halo
at the distance of the satellite, and an {\em evaporation} effect
induced by the rapidly varying tidal force near pericenter radii for
satellites moving on an eccentric orbit. In the latter case we speak
of tidal shocks -- short impulses are imparted to bound particles
within the satellite, heating the system and causing its dissolution.

\subsection{The tidal truncation} 
A tidally limited satellite is truncated at its tidal radius $R_{\rm
s,tid}$, which, loosely 
speaking, corresponds to the distance (relative to the satellite
center) at which the mean density of the
satellite is of the order of the mean density of the 
hosting halo, at the satellite position $r$:
\beq 
\bar\rho_s(R_{\rm s,tid}) \approx  \bar \rho_{\rm h}(r)\;.  
\label{eq:tidcut}
\eeq 

The evaluation of the tidal radius requires a relation between $R_{\rm
s,tid}$ and $r$ which is customarily derived from the force equivalence
between internal gravity and external tides leading to the implicit
equation:
\beq 
\label{eq:rtidtds}	
R_{\rm s,tid}=r\, \left \{ 
 { M_{\rm s}(R_{\rm s,tid}) \over \left 
(2-\partial \ln M_{\rm h}/\partial \ln r \right) 
M_{\rm h}(r)} \right\}^{1/3} 
\eeq
(Tormen, Diaferio \& Syer 1998).  The mass tidally lost, $\Delta
M_{\rm s,tid}$ is thus computed subtracting spherical shell above
$R_{\rm s,tid}$, using equation~(\ref{eq:nfwmass}) .  While strictly
valid for a satellite moving on a circular orbit (where the combined
potential over the system is static in the satellite frame) $R_{\rm
s,tid}$ gives, if evaluated at every single point $r$ (Binney \&
Tremain 1987), an approximate expression for the instantaneous tidal
radius in the case of non circular motion.
This implies that, on stable orbits, $\Delta M_{\rm s,tid}$ is maximum
at the first pericenter passage; the mass of the satellite would then
remain constant. In Fig. (3) we give the residual mass
after instantaneous tidal cut, as a function
of circularity, as computed using equation~(\ref{eq:rtidtds}).

Tidal stripping however does not occur instantaneously, and,  following TB
suggestion, we model mass loss, over a few orbital periods, adopting
the expression
\beq {dM \over dt} \simeq {\Delta M_{\rm s,tid} (t) \over 2 \pi /\omega(t)}
\label{eq:tidal_cut}
\eeq
where $\omega(t)$ is the instantaneous orbital angular velocity.  This
is compared with results from numerical simulations.
Fig.~(\ref{fig:dis}) gives the satellite mass as a function of time
for a selected run. We find that mass loss by tidal cut, as described
by equation~(\ref{eq:tidal_cut}), reproduces the result of our N-body
simulation only in the early phase: the satellite loses mass at a
peace larger than predicted by equation (\ref{eq:tidal_cut}) (we refer
to the dashed line of Fig.~[\ref{fig:dis}]).  We believe that this is
due to the action of tidal shocks (and not a numerical artifact).


The number of particles in the N-body simulations is chosen in order to avoid as much as
possible numerical two-body relaxation which could increase,
artificially, the overall evaporation rate (Gnedin \& Ostriker 1999;
Moore, Katz \& Lake 1996) .  Numerical relaxation disperses satellite
particles over a timescale related to the number of particles $N$
\beq
\label{eq:trh:18}
t_{\rm rh}=0.138 \, {M_{\rm s}^{0.5} R_{\rm hm}^{1.5} \over G^{0.5} m_*
\ln(0.4N)}\;.
\eeq
where $ R_{\rm hm}$ is the half mass radius and $m_*$ is the particle
mass, and $M_s=m_*N$.  As shown in Table~(1) the initial relaxation
time is $\sim$ 100 Gyr and remains longer than 10 Gyr as mass loss
continues.  This is an indication that numerical two-body relaxation
is unimportant. We thus proceed on modeling mass loss with the
inclusion of tidal shocks.  In the next Section we estimate the
shock-induced evaporation time and show that, for NFW satellites, it
can be shorter than cosmic age.

\subsection{Heating \& evaporation}

The description of the dynamical evolution of a satellite must
include also  tidal heating due to compressive tidal shocks.

The theory of shock heating was developed by Ostriker et al. (1972)
and Spitzer (1987) to model the evolution of globular clusters. Recent
works by Gnedin \& Ostriker (1997) and GHO extend this theory also to
tidal perturbation on satellites moving on eccentric orbits inside an
extended mass distribution.  We will use the GHO model to treat tidal
shocks on satellites orbiting specifically inside a NFW halo. At each
pericenter passage satellites cross the denser regions of the main
halo: the rapidly varying tidal force induces a gravitational shock
inside the satellite. The shock increases the velocities of satellite
particles, and reduces the satellite binding energy. As a result the
satellite expands.

The amount of heating is a function of the orbital parameters and of
the concentration of the main halo:
\beq
\langle \Delta E \rangle =
{\cal F}\left(c_h,x_{\rm c} [E],\cir\right) A(x_\tau)
\cdot R_{\rm s}^2 \;,
\eeq
The shock is more intense in the outer layers, as it depends also on
the satellite radius $R_{\rm s}$ (see Appendix A for the details of
the calculation). As suggested by GHO, we use an adiabatic correction
$A(x_\tau)=(1+x_\tau^2)^{\gamma}$ with $\gamma=-5/2$ (see also
Weinberg 1995) . Here $x_{\tau} \equiv \omega \tau$ is the adiabatic
parameter, $\tau$ is the duration of the shock and
$\omega=\sigma_s(R_{\rm s})/R_{\rm s}$ where $\sigma_s$ is the
velocity dispersion of particles in the satellite at radius $R_{\rm
s}$.  The value of $\tau$ is related to the pericenter crossing time:
we assume $\tau = 0.5/\omega_{\rm per}$, where $\omega_{\rm per}$ is
the orbital angular velocity at pericenter distance. The adiabatic
correction accounts for the fact that the susceptibility of a system
to the tidal shocks will also depend on its internal dynamics - when
the internal orbital time is very short a particle in the satellite
will receive two opposite tidal kicks of nearly the same magnitude and
the net effect will be small (e.g.  Binney \& Tremaine 1987).

\begin{figure}
\centerline{\psfig{figure=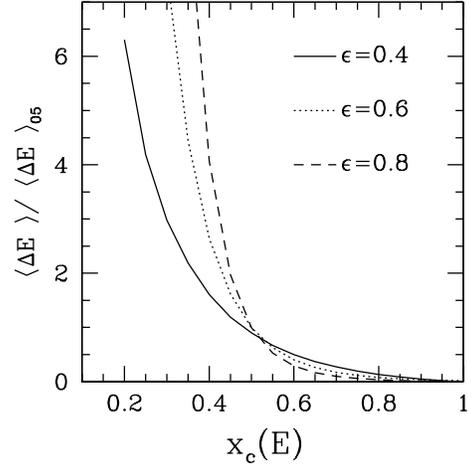,width=7.0cm}}
\caption[]{ The amount of  shock heating as function of the
circular radius $x_{\rm c} (E)$.  $\langle \Delta E \rangle$ is
normalized to the value of  $\langle \Delta E \rangle$ 
 when $x_{\rm c} (E)=0.5$ and $\cir=0.6$.  We consider three different values of
the circularity: $\epsilon=0.4$ (solid line), $\epsilon=0.6$ (dotted
line) and $\epsilon=0.8$ (dashed line).  The satellite and halo
concentration are chosen such as $c_{\rm s}/\c=2$.  }
\label{fig:DeltarcE}
\end{figure}

\begin{table*}
\begin{minipage}{100mm}
\caption{The characteristic timescales}
\label{tab:time}
\centerline{
\begin{tabular}{lcccrr}
 Model & $c_{\rm s}/\c$ & $t_{\rm sh}$[Gyr] & $P_{\rm orb}
 $[Gyr] & $t_{\rm rh}$[Gyr] \\
\hline \\
Low concentration &  &  &   &  &  \\
$\epsilon = 0.7$ \quad $x_{\rm c} =0.5$  &  0.5 &  12.6 & 4.7 & 176.4 \\
$\epsilon = 0.5$ \quad $x_{\rm c} =0.3$  &  0.5 &   0.7 & 2.6 & 173.6 \\
Intermediate concentration &  &  &   &  &  \\  		             
$\epsilon = 0.7$ \quad $x_{\rm c} =0.5$  &  1   &  93.6 & 4.7 & 119.2 \\
$\epsilon = 0.5$ \quad $x_{\rm c} =0.3$  &  1   &  2.00 & 2.6 & 112.6 \\
High concentration &  &   &  &  \\	       		             
$\epsilon = 0.7$ \quad $x_{\rm c} =0.5$  &  2   &  130  & 4.7 &  80.7 \\
$\epsilon = 0.5$ \quad $x_{\rm c} =0.3$  &  2   &  6.6  & 2.6 &  73.8 \\
\hline \\
\end{tabular}
}
\end{minipage}
\end{table*}

\begin{figure}
\centerline{\psfig{figure=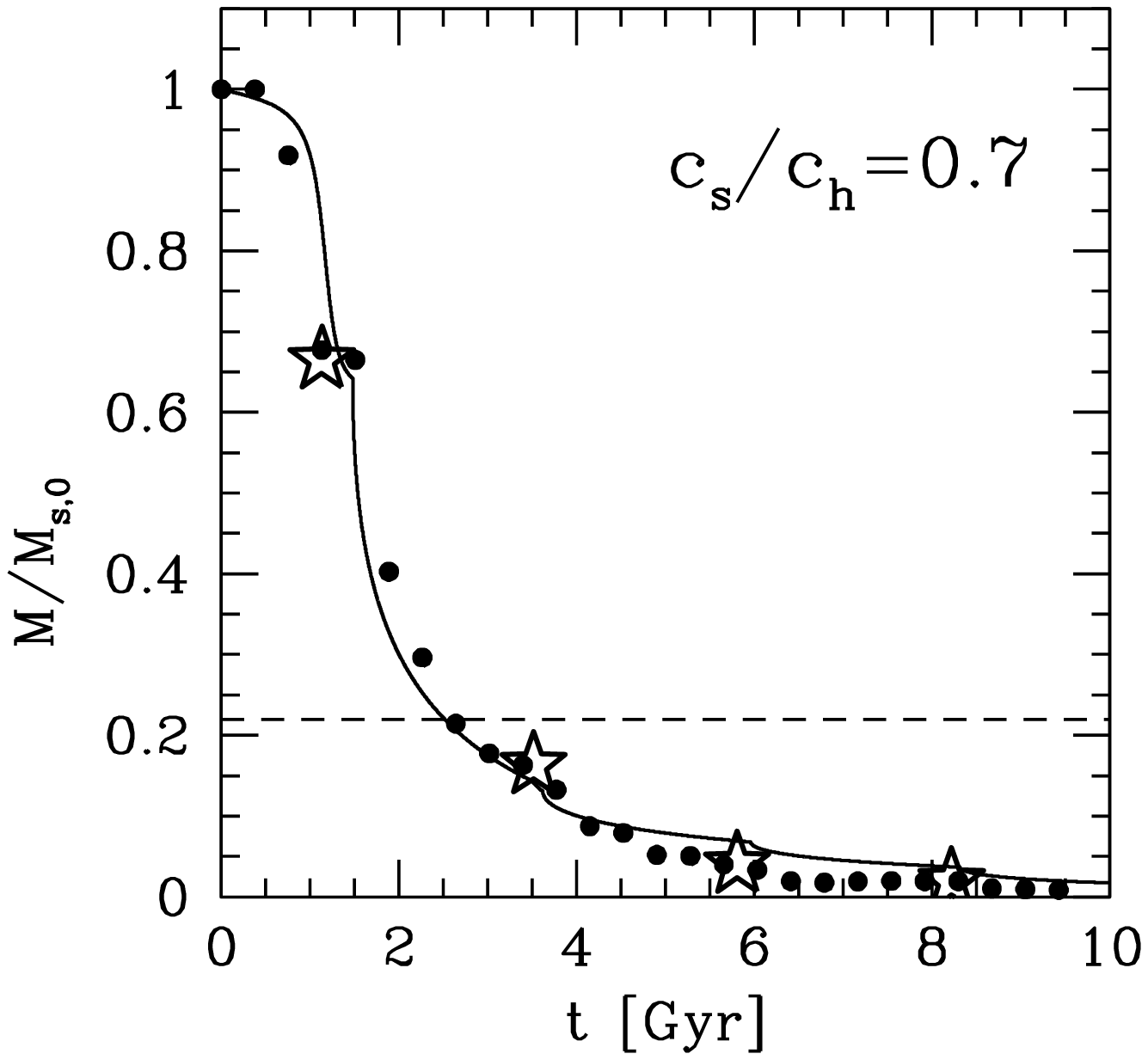,width=7.0cm}}
\centerline{\psfig{figure=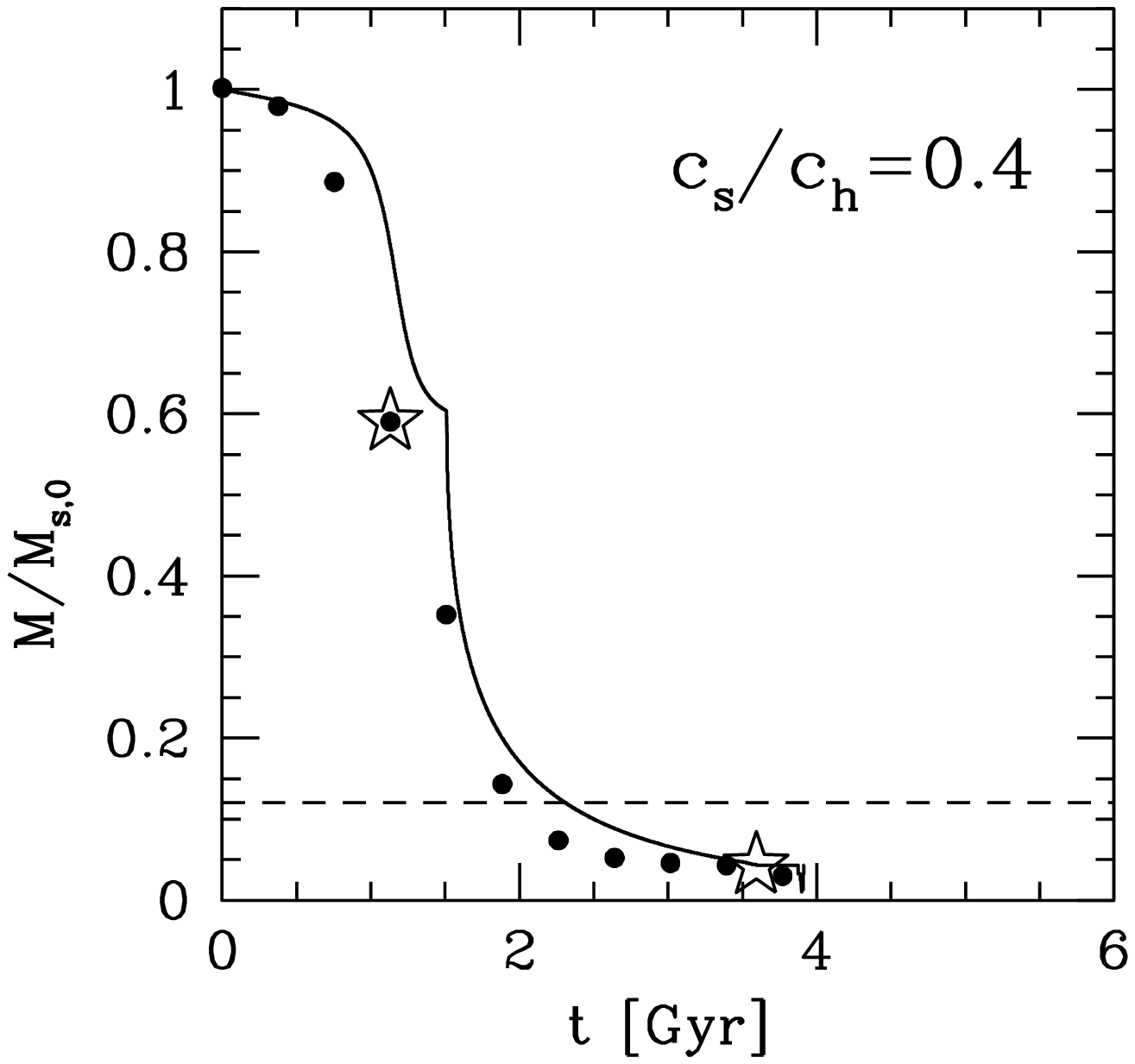,width=7.0cm}}
\caption[]{Bound mass in units of the initial mass
as a function of time, for a satellite moving
on a stable orbit.  The orbital parameters are  $\epsilon =
0.65$ and $x_{\rm c} (E)=0.5$. The concentration ratio is $c_{\rm
s}/\c=0.7$ (top panel) and $c_{\rm s}/\c=0.4$ (bottom panel).  The
symbols are the N-body data and the solid line the
semi-analytical model. Stars  
identify each pericenter passage.  The dashed line is the bound 
mass that would remain if we apply only the tidal cut using 
equation~(\ref{eq:tidal_cut})
over a few orbital periods.}
\label{fig:dis}
\end{figure}

We introduce a characteristic shock timescale computed, after each
pericenter passage, as
\beq
t_{\rm sh'}=\frac {P_{\rm orb}}{2} 
\cdot {E_0 \over \langle \Delta E_{\rm hm}\rangle} \;, 
\eeq 
where $E_0=0.25 G M_{\rm s,per}/R_{\rm s,hm}$ is the binding energy of
the tidally truncated satellite of mass $M_{\rm s,per}$ evaluated
according to equation~(\ref{eq:tidal_cut}) at the time of pericentric
passage.  Both $E_0$ and $\langle \Delta E_{\rm hm}\rangle$ are
evaluated at the half mass radius $R_{\rm s,hm}$ which is a function
of the satellite concentration. A second order energy change due to
shock heating is responsible for increasing the internal velocity
dispersion and allows additional particles to leave the satellite. To
account of this second order perturbation we assume that $t_{\rm
sh}=0.43 t_{\rm sh'}$ (Gnedin, Lee
\& Ostriker 1999; hereafter GLO).
Table~(\ref{tab:time}) shows the shock time for the satellite
modeled at first pericenter passage. The number of pericenter
passages roughly necessary to unbind the satellite is $t_{\rm sh}/P_{\rm
orb}$.
Lastly we notice that $\langle \Delta E \rangle$ increases
linearly with the halo concentration $\c$, because  in highly
concentrated haloes the gradient of the gravitational force
is steeper.

The amount of heating is also a function of the orbital parameters; in
Fig.~(\ref{fig:DeltarcE}) we study the energy gain as a function of
$x_{\rm c} (E)$ for different values of the circularity. The fast
growth of $\langle \Delta E\rangle$  for small values of
$x_{\rm c} (E)$ confirms that shocks on radial orbits are more
intense:  a satellite moving on a circular orbit  is not subject to
any heating.

\subsection{Modeling the mass loss} 

Tidal shocks are events leading to the escape of
particles. To model the induced mass loss, we introduce the so called
escape probability function $\xi_{\rm sh},$ 
analog of $\xi_{\rm e}$ customarily used to
describe globular cluster evaporation by two-body relaxation
processes (Spitzer 1987).

Mass loss can be predicted using the dimensionless rate of
escape
\beq
\label{eq:evap}
\xi_{\rm e} \equiv -{t_{\rm rh} \over M(t)}{d\,M \over d\,t}\;.
\eeq
Similarly here we define 
\beq
\label{eq:evap.sh}
\xi_{\rm sh} \approx -{t_{\rm sh} \over M(t)}{d\,M \over d\,t}\;,
\eeq
For the case of escape by two-body relaxation  $\xi_{\rm sh}$
is a constant known to vary from $7.4\cdot 10^{-3}$ for an isolated
halo to $0.045$ for a tidally truncated halo (Spitzer 1987).  On the
contrary, when tidal shocks are present and dominate, the escape
probability becomes a function of time; $\xi_{\rm sh}$ peaks just
after each pericenter passage (GLO), rapidly decreasing until the next shock
event.  $\xi_{\rm sh}$ is then a periodic function of period $P_{\rm
orb}$ and we find that it can be fitted using both simulations and
results by GLH as:
\beq
\label{eq:xitime}
\xi_{\rm sh}(t) \propto 
\left({t - t_{\rm per} \over t_{\rm tr}}\right)^{-0.5} 
\exp - \left( {t-t_{\rm per} \over t_{\rm tr}}\right )^{0.5}\;,
\eeq
where $t_{\rm tr}\simeq 13 \, t_{\rm sh}$ and $t_{\rm per}$ is the
pericenter time.  The shock escape probability is equal to unity at
$t=t_{\rm per}+t_{\rm dyn}$ where $t_{\rm dyn}$ is the dynamical time
of the satellite. 
The shock time must be evaluated at each pericenter
passage as it varies according to the current  orbit 
and mass of the satellite.

The orbital timescale is sometime shorter that the shock timescale
so that the satellite suffers. 
If it becomes unbound, further dispersal of the last
particles occurs on the crossing time of the damaged system.

\subsection{Testing the model for a live satellite}
The dynamical evolution of a satellite is described using a
semi-analytical code which accounts of both dynamical 
friction and mass loss. 
In this context, we
use the expression of the drag force 
as given in equation~(\ref{eq:nfworbeq}), 
since it is much faster, and in close match with TLR 
(see Section 3). At each
time step we upgrade the satellite mass according to
equation~(\ref{eq:tidal_cut}) and equation~(\ref{eq:evap.sh}).
To test the ability of our code to follow the 
evolution of an NFW satellite, we compare the results with those
derived from a selected set of N-body simulations.

\subsubsection{Tidal perturbations on stable orbits}
To isolate and study the effect of a pure tidal perturbation we explore
the dynamical evolution of a low concentration satellite on an
unperturbed orbit.  In this case,
the heating by tidal shocks varies solely as a consequence of the
progressive reduction of the satellite half-mass radius. For this reason we
expect a progressive reduction of the shock destructive power as time
passes.
\begin{figure*}
\centerline{\psfig{figure=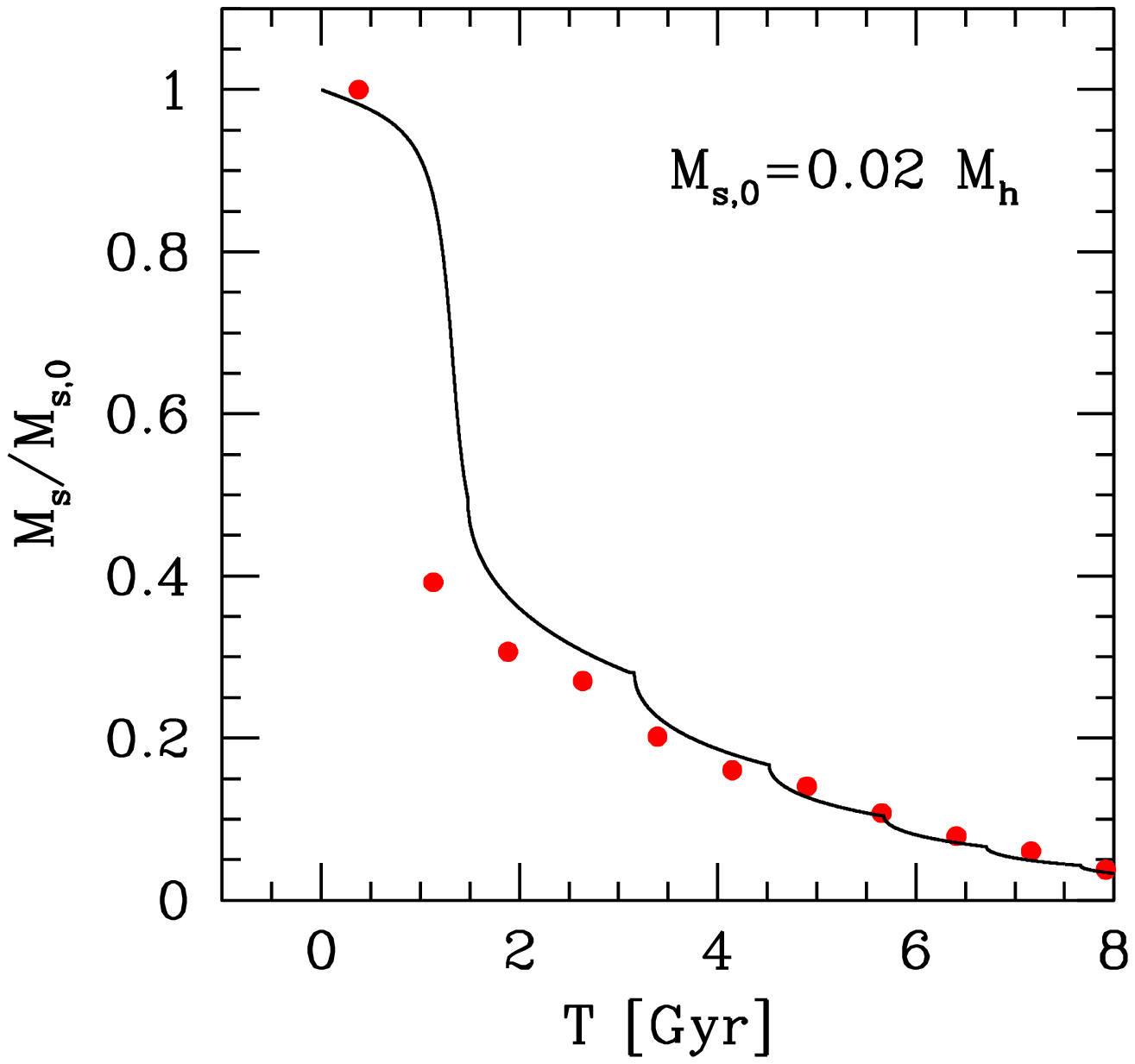,width=7.0cm}\psfig{figure=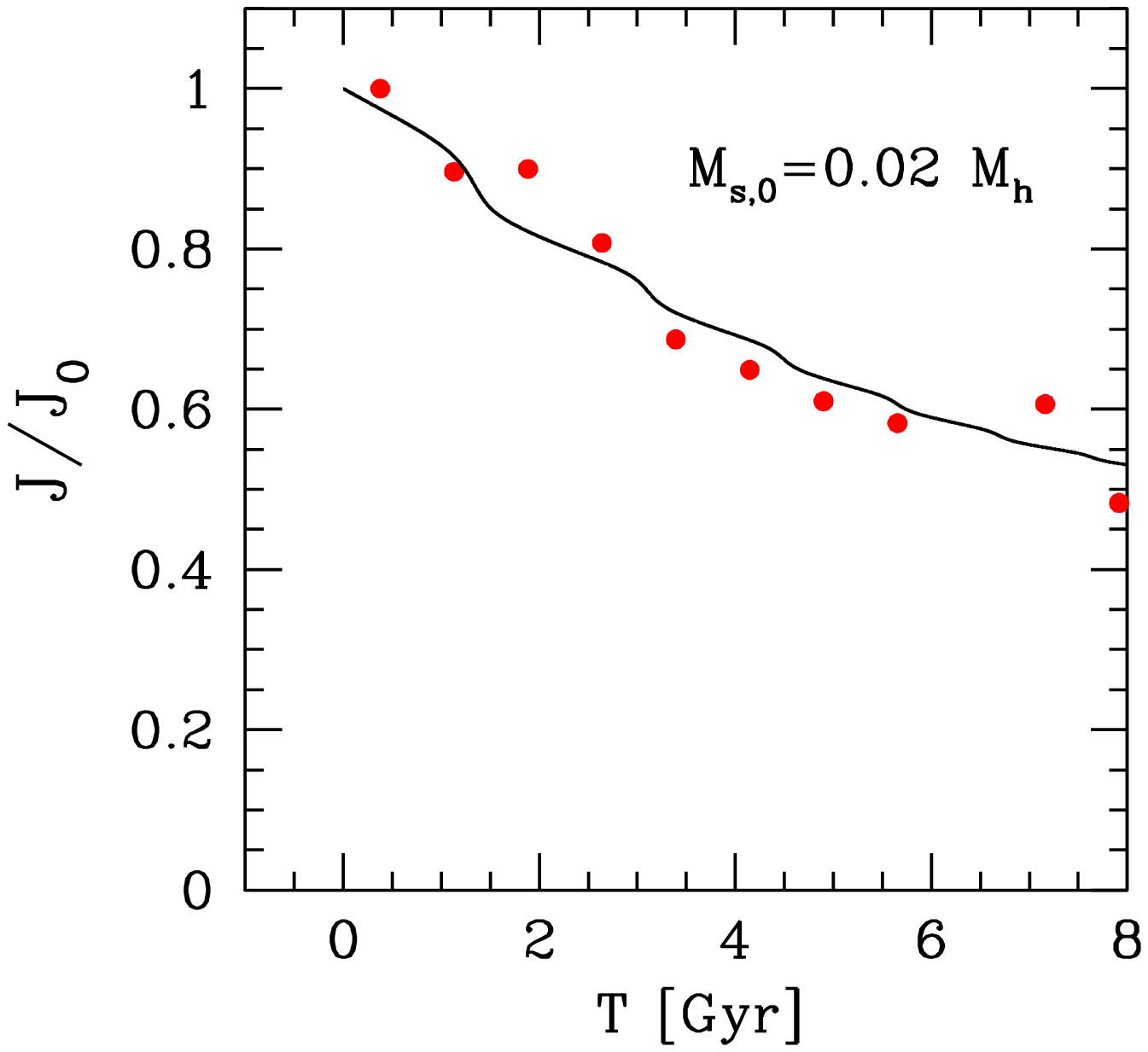,width=7.0cm}}
\centerline{\psfig{figure=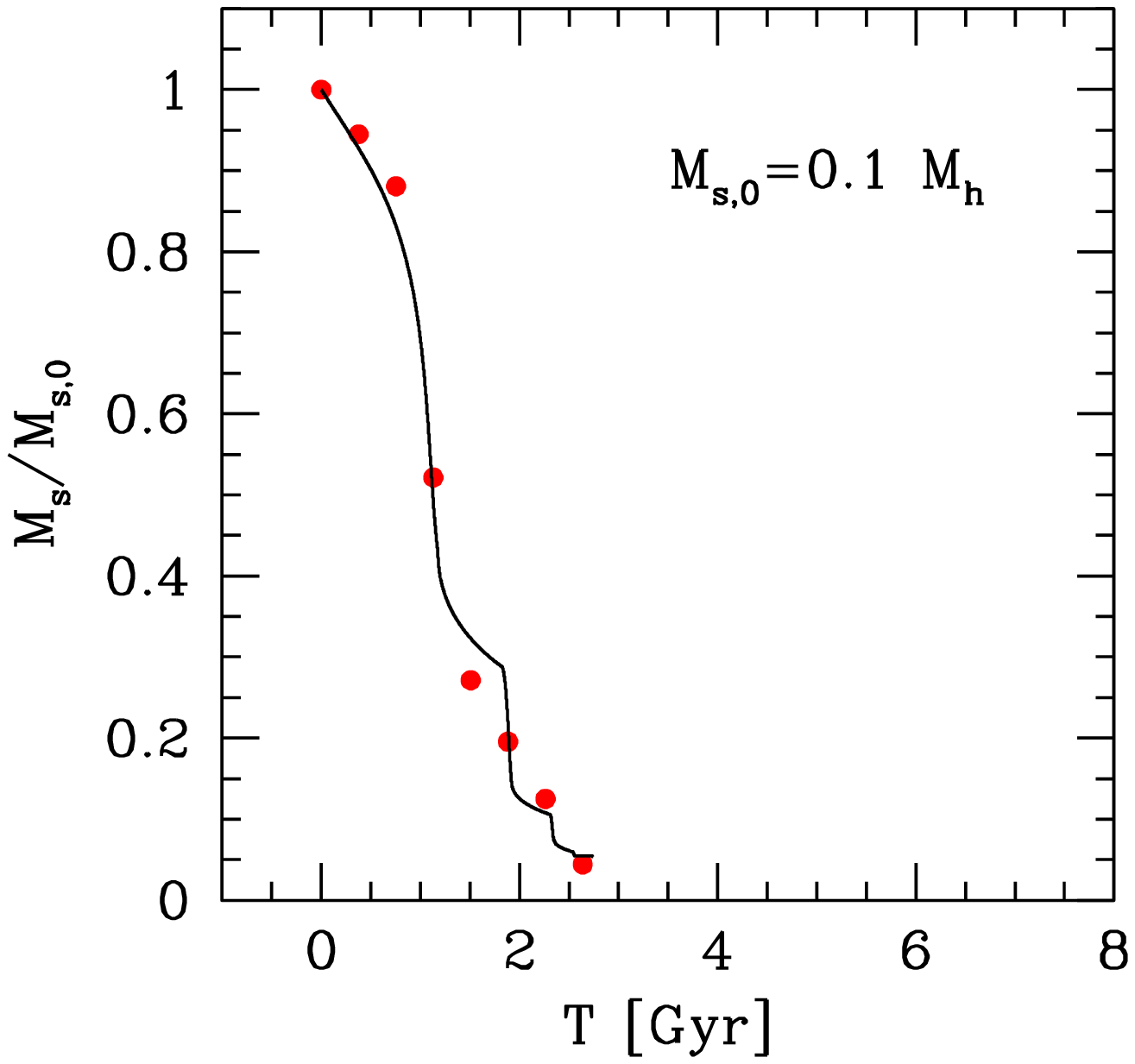,width=7.0cm}\psfig{figure=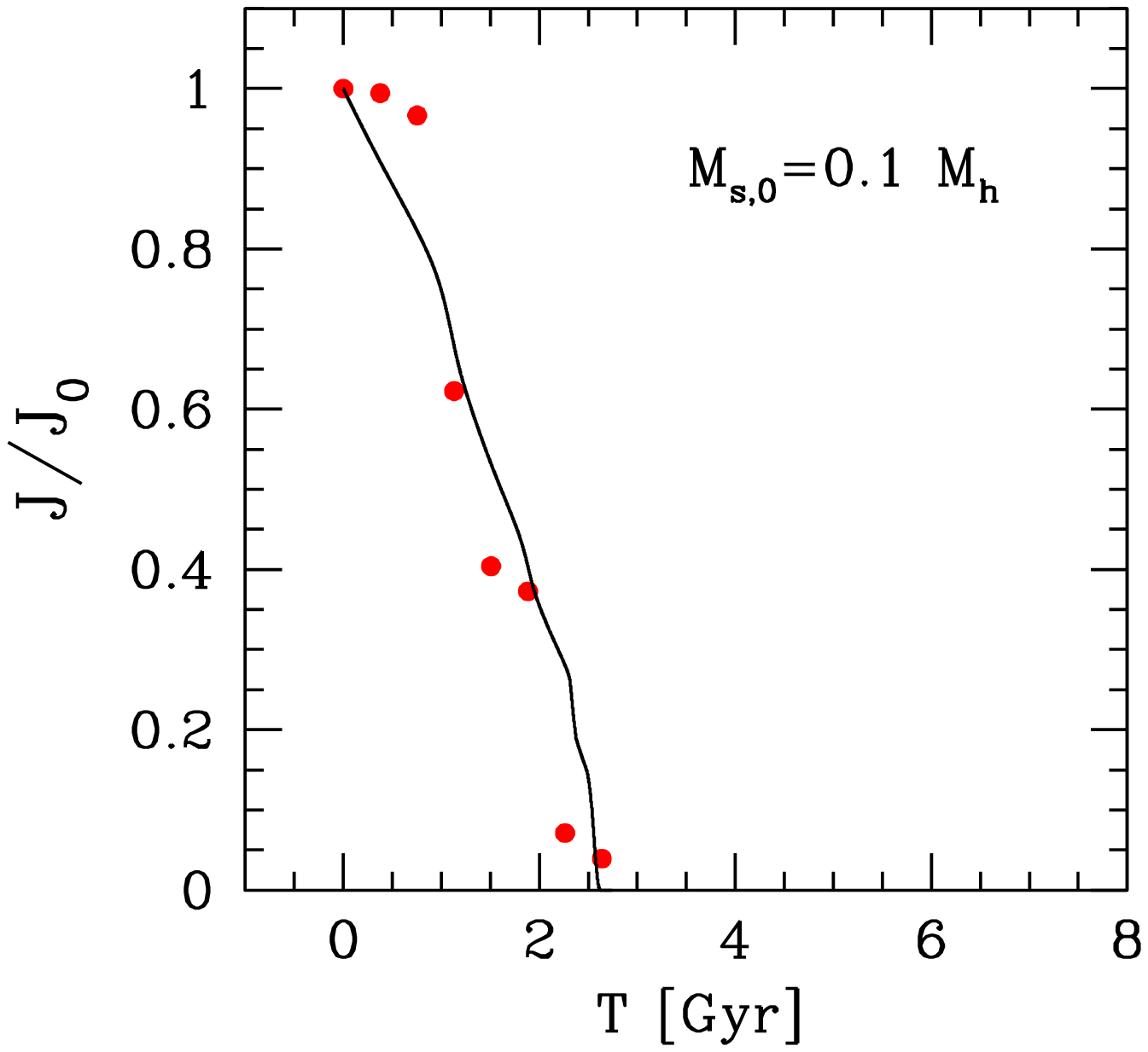,width=7.0cm}}
\caption[]{
We plot the time evolution of the mass,
in units of the initial mass (left panels), and orbital angular
momentum scaled to the initial one (right panels). The adopted values are:
$M_{\rm s,0}=0.1 \, M_{\rm h}$ and $M_{\rm s,0}=0.02 \, M_{\rm h}$.
The concentration ratio and orbital parameters are
$c_{\rm s}/\c=2$, $\cir = 0.7$ and
$x_{\rm c} (E)=0.5$, respectively. Points are from N-body runs and solid line
from our 
semi-analytical model.} \label{fig:dfms}
\end{figure*}
In Fig.~(\ref{fig:dis}) we show the evolution of a satellite with mass
$M_{\rm s,0}=0.01 M_{\rm h}$; the orbital parameters are chosen to
reflect a typical cosmological orbit: $\cir = 0.65$ and $x_{\rm
c}(E)=0.5$.
The bottom panel shows a low $c_{\rm s}/\c$ satellite,
disrupted after the second passage to pericenter.
The top panel shows the evolution of a higher $c_{\rm s}/\c$
satellite surviving for more than 12 Gyr, despite having lost  more
than the 99 per cent of its mass.
Until the first pericenter passage only the tidal cut accounts for the
mass loss. The good agreement between the simulation and the code
before the first pericenter passage suggests that the TB recipe is
accurate enough to reproduce the mass loss before (or in absence of)
the shock heating.

\subsubsection{The combined effect of dynamical friction 
and tidal stripping} 

The dynamical evolution of a satellite is driven by the combined
effect of dynamical friction that drives the satellite to the center
of the main halo, and the tidal perturbation which reduces its mass.
The two processes are intimately connected as the drag force is
strongly related to the mass and size of the satellite.

In Fig.~(\ref{fig:dfms}) we compare the semi-analytical model with the
the results of N-body simulations for satellites with $c_{\rm
s}/c_{\rm h}=2$. The initial orbital parameters are $\epsilon=0.7$,
and $x_{\rm c} (E)=0.5$.  We study two different cases: a light
satellite of mass $M_{\rm s,0}=0.02\,M_{\rm h}$ and a massive one with
$M_{\rm s}=0.1\,M_{\rm h}$.  The mass loss rate and the orbital
evolution are well reproduced in both cases.
The massive satellite loses mass during evolution, yet a
core of bound particles survives having 5\% of its initial mass but
sinks to the center merging with the main halo in 3 orbital
periods. On the contrary, the light satellite loses 99\% of
its mass but a bound core remains which moves on an inner orbit
stable against dynamical friction, following mass loss.

\section{The fate of satellites }

\subsection {The recipe}

\begin{figure}
\centerline{\psfig{figure=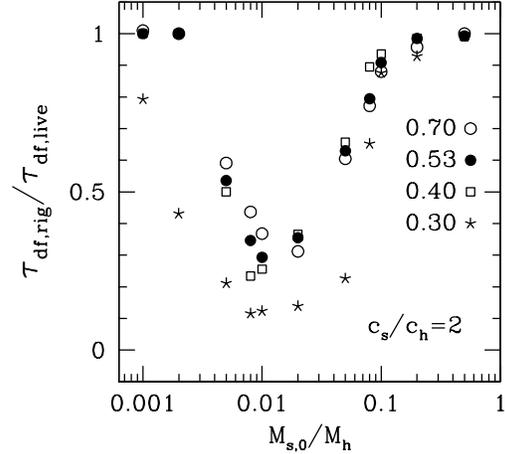,width=7.0cm}}
\caption[]{
We have plotted the ratio of the dynamical friction time of
a non-deformable satellites to the same time for a
live satellite of equal mass, initially, 
as function of $M_{\rm s,0}/M_{\rm h}$. The concentration ratio is 
$c_{\rm s}/\c=2,$ and  
$x_{\rm c} (E)=0.5.$ 
We vary the circularity which is labeled with different symbols.
We notice that symbols for $\cir=0.3$ refer to a different
orbital energy: $x_{\rm c} (E)=0.3.$}
\label{fig:tdtl}
\end{figure}

We now use our semi-analytical model to quantify how mass loss affects the orbital
decay.  In Fig.~(\ref{fig:tdtl}) we give as a function of
$M_{\rm s,0}/M_{\rm h},$ the ratio of the dynamical
friction time of a rigid satellite $\tau_{\rm df,rig}$ to the
same time  for a homologous live satellite $\tau_{\rm df,live}$
\footnote{We use highly concentrated satellites ($c_{\rm s}/c_{\rm
h}=2$) that are not rapidly disrupted by tidal interactions.}.
In taking the ratio we mainly quantify the importance of
mass loss in affecting the lifetime of a satellite.  Fig.~(\ref{fig:tdtl}) 
shows that massive satellites ($M_{\rm s,0}/M_{\rm h}>0.1$)
sink to the center of the main halo on a timescale  $\tau_{\rm df,
life}\sim \tau_{\rm df,rig}$ as if they were rigid.
In the mass range $M_{\rm s,0}\simeq0.005-0.1 M_{\rm h}$ the satellites 
sink toward the center of the main halo by dynamical friction and so  lose mass 
efficiently. Accordingly, the dynamical friction  time $\tau_{\rm df,live}$ 
is a few  to several times {\it longer} than $\tau_{\rm df,rig}.$ 
At lower masses, $M_{\rm s,0} \sim < 0.005 \, M_{\rm h},$
orbits are less perturbed by friction and mass stripping
becomes less important. 
Thus the ratio $\tau_{\rm df,rig} / \tau_{\rm df, life}$  starts to
rise again. 


We now estimate the dynamical friction timescale for a live satellite 
in three different regimes. 

\noindent For massive satellites, $M_{\rm s,0} \ge 0.1 \;M_{\rm h}$,
the dynamical friction time is not affected by the mass loss so
\beq 
\label{eq:fittau01}
\tau_{\rm df,live} \sim \tau_{\rm df,rig} \simeq 
0.5 \:{R^2_{\rm h} V_{\rm h}\over G M_{\rm s}}\: 
{\cal A}_{\rm rig} \left[{M_{\rm s,0} \over  M_{\rm h}}, \c,x_{\rm c} (E) \right] \;,
\eeq
where:
\beq
{\cal A}_{\rm rig} \left[{M_{\rm s,0} \over M_{\rm h}}, \c, x_{\rm c} (E)
\right] = f(\c)\, { x_{\rm c}^{1.97} (E) \over 
\ln \left( 1+{ M_{\rm h} / M_{\rm s,0} } \right) }\;,
\eeq
and $f(\c)$ is given by equation~(\ref{eq:fdic}).
In this case, the dynamical friction time depends weakly on  
$\cir$;  the exponent $\alpha \approx 0$, as indicated by equation (\ref{eq:alfa}). 
\noindent For ${0.007 \; M_{\rm h} < M_{\rm s,0} < 0.08\; M_{\rm h}}$,
we provide a  fit of the form:
\beq
\label{eq:fittau001}
\tau_{\rm df,live} \sim  {R_{\rm h}^2 \; V_{\rm h} \over {G M_{\rm s,0}}}
{\cal A}_{\rm live} \left[{M_{\rm s,0} \over M_{\rm h}}, 
\frac{c_{\rm s}}{\c} ,x_{\rm c} (E), \epsilon \right]\;.
\eeq
For $\cir=1$ 
$$
{\cal A}_{\rm live} \left({M_{\rm s,0} \over M_{\rm h}}, \frac{c_{\rm s}}{\c},
x_{\rm c}, \epsilon=1 \right)= \left[{0.25 \over (c_{\rm s}/\c)^6}-0.07\, \frac{c_{\rm s}}{\c}+1.123 \right]
$$
\beq
~~~~~~~~~~~~~~~\times
\left[ B(x_{\rm c})\left({M_{\rm s,0} \over M_{\rm h}}\right)^{0.12}
+C(x_{\rm c})\left({M_{\rm s,0} \over M_{\rm h}}\right)^2\right] 
\;,
\label{eq:cal.a}
\eeq
where
\beq
B(x_{\rm c})=-0.050+0.335\, x_{\rm c}+0.328\,x_{\rm c}^2 \;, 
\eeq
\beq 
C(x_{\rm c}) = 2.151-14.176\, x_{\rm c}+27.383\,x_{\rm c}^2
\;.
\eeq
This fit reproduces the semi-analytic estimate of the
decay time of a live satellite on circular orbits with an error of
$\gse 9$ per cent, for $0.2 < x_{\rm c}(E) < 1$.
\begin{figure}
\centerline{\psfig{figure=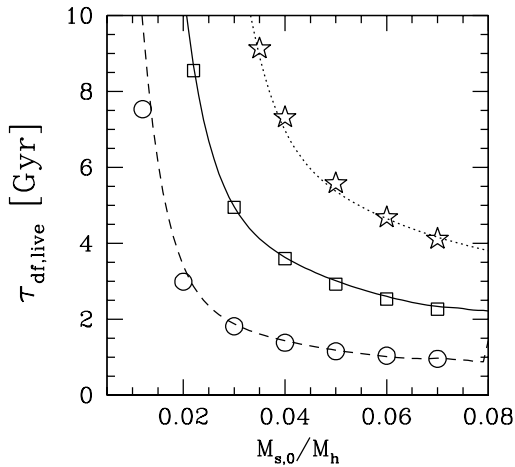,width=7.0cm}}
\caption[]{
The dynamical friction time of life satellites
As function of the satellite initial mass.
 Symbols are the  results obtained from the semi-analytic code, lines  are the prescription of the fitting formulae (from equation~[27]   to equation~[32]).
We set the circularity $\cir=0.7$ and the relative concentration
$c_{\rm s}/\c=2$ and we vary the orbital energy:  $x_{\rm c} (E)=0.5$ (solid line and squares),  $x_{\rm c} (E)=0.7$ (dotted line and stars), $x_{\rm c} (E)=0.3$ (dashed line and circles). 
}
\label{fig:fitlive2}
\end{figure}


For eccentric orbits we find that:
\bea
\nonumber
{\cal A}_{\rm live} \left({M_{\rm s,0} \over M_{\rm h}},
\frac{c_{\rm s}}{\c}, x_{\rm c}, \epsilon \right) =
{\cal A}_{\rm live} \left({M_{\rm s,0} \over M_{\rm h}}, 
\frac{c_{\rm s}}{\c}, x_{\rm c}, \epsilon=1\right) \, \\
\times
\left[ 0.4+{\cal Q}\left({M_{\rm s,0} \over M_{\rm h}}, x_{\rm c}\right)
\times (\cir-0.2)
\right]
\;,
\eea
where
\bea
\nonumber
{\cal Q}\left({M_{\rm s,0} \over M_{\rm h}}, x_{\rm c}\right)=
0.9+ 10^8\left(12.84+3.04\, x_{\rm c}-23.4 \, x^2_{\rm c} \right) \\
\times
\left({M_{\rm s,0} \over M_{\rm h}}-{0.0077 \over 1-1.08 x_{\rm c}}-0.0362  \right)^6
\eea
This formula 
holds when $ 0.3<x_{\rm c}<0.9$ and $0.3<\cir<0.8$ and reproduces the
semi-analytical data within and error of $15\%$.
 
Interestingly, we notice that for an eccentric orbit the decay time can be
longer that the \df\ time on the circular orbit with the same initial
orbital energy, since mass loss on eccentric orbit is higher because of
the tidal shock. 

For $\cir>0.8,$ ${\cal {A}}_{\rm live}$ we use 
the interpolation of equations (31) and 
(28). If ${0.08 \; M_{\rm h} < M_{\rm s,0} < 0.1\; M_{\rm h}}$, we suggest to linear interpolate 
equations (25) and (27).

Satellites $M_{\rm s,0} \lse 0.007 \;M_{\rm h}$, evolve on
slightly perturbed orbits, the dynamical friction timescale in this
case is al least two times longer than for the rigid satellite (CMG99 similarly
found an increase of a factor $e$ for $M_{\rm s,0}=0.02M_{\rm h}$).
We suggest to use equation  (12) for $\tau_{\rm {df,life}},$ in this mass
range, increased by a factor $\sim 2.73$, together with equation (15) for the exponent
$\alpha,$ to account for the correction to circularity. For very light satellites 
($M_{\rm s,0}<10^{-4} M_{\rm h}$) equation~(\ref{eq:taualpha}) holds.

\begin{figure*}
\centerline{\psfig{figure=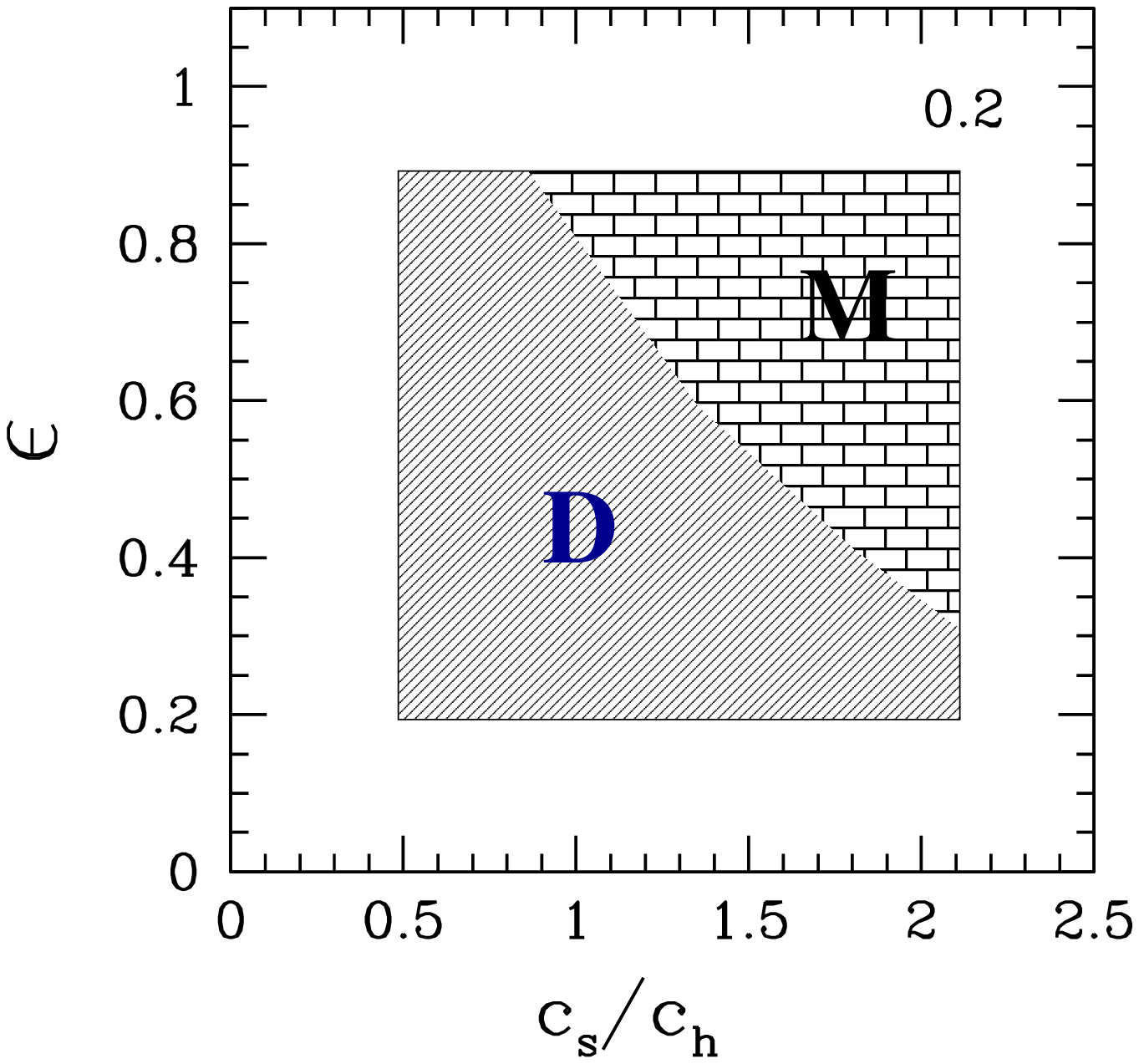,width=6.0cm}\psfig{figure=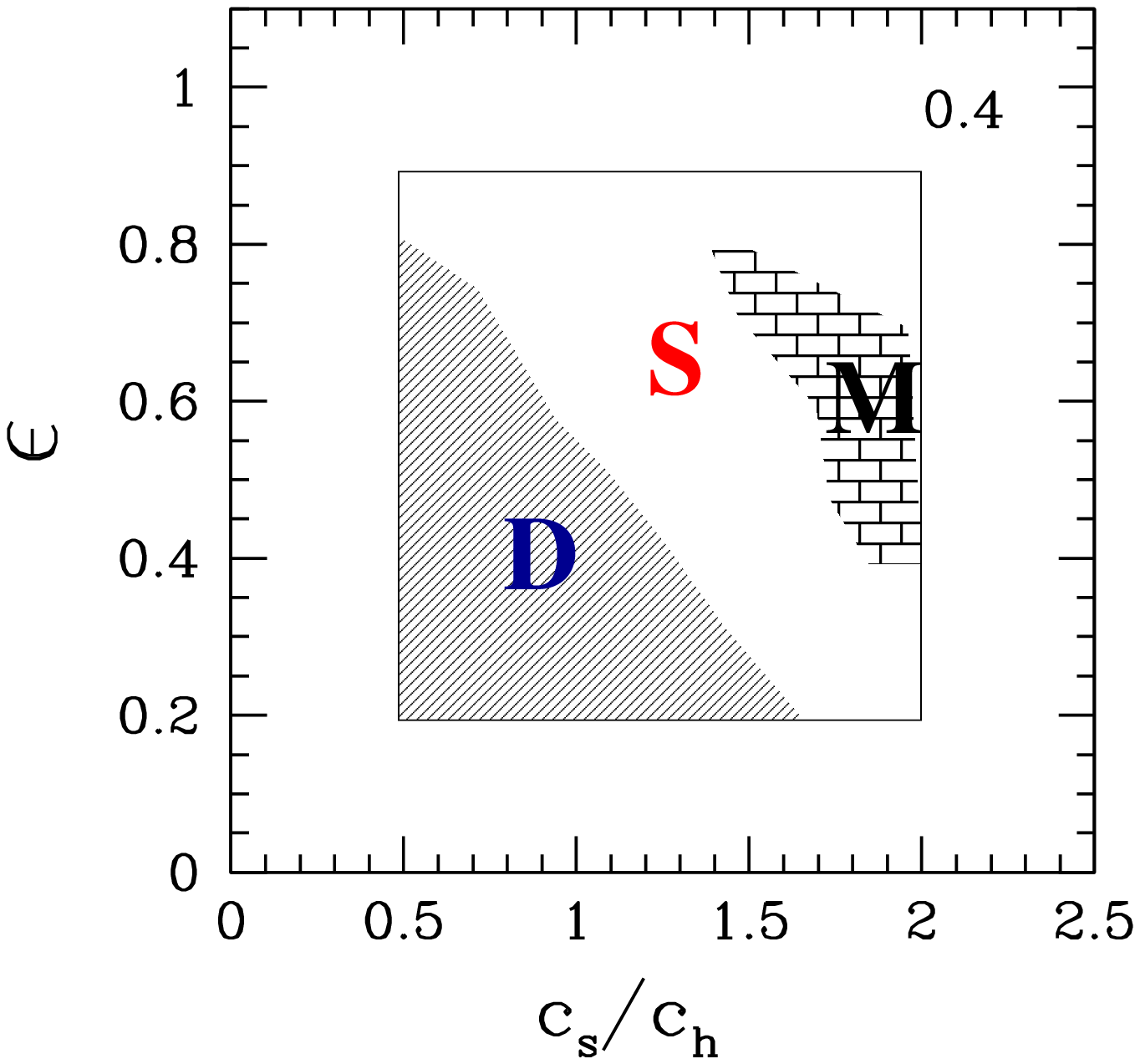,width=6.0cm}\psfig{figure=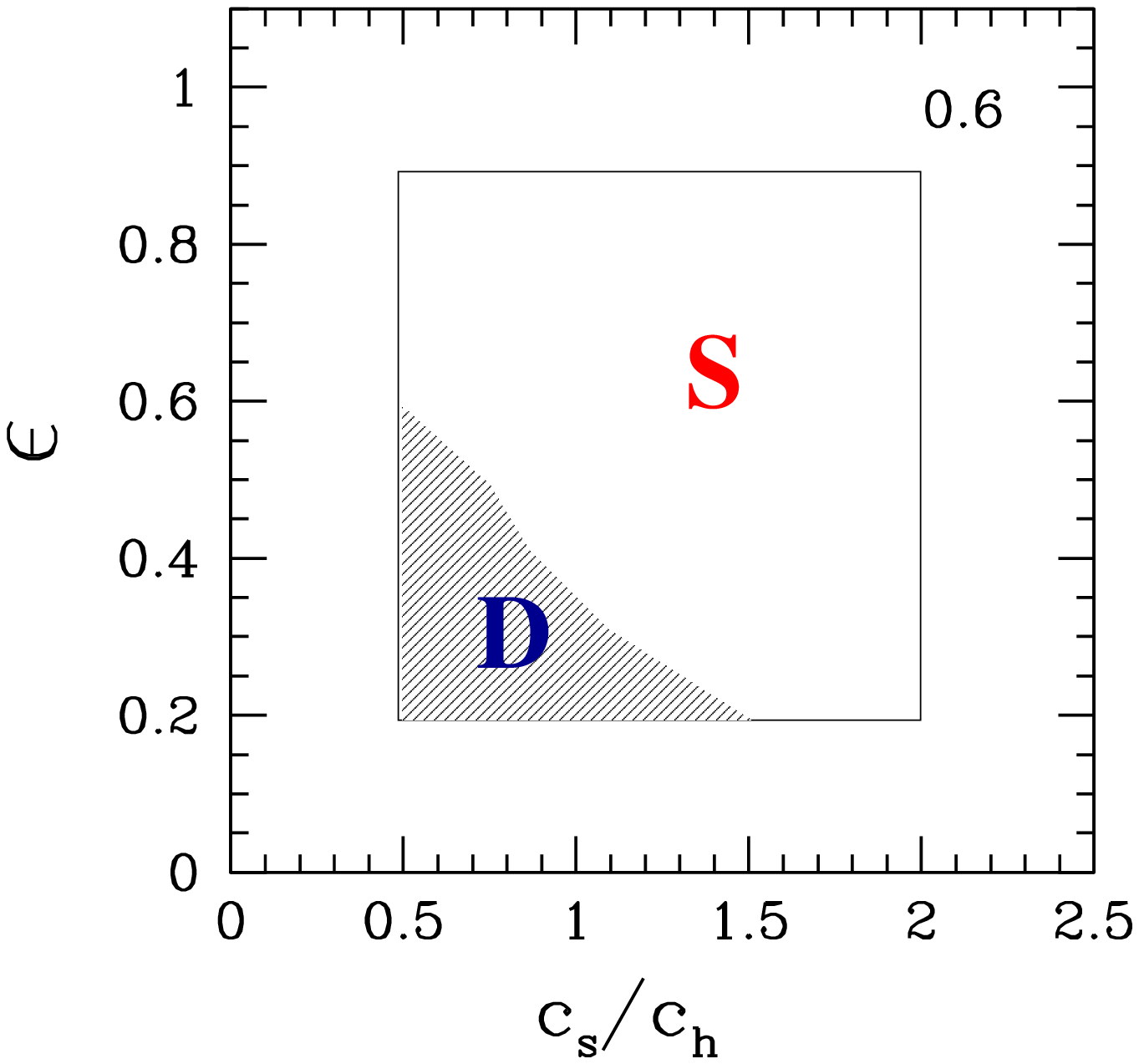,width=6.0cm}}
\caption[]{The life diagram of a satellite with  $M_{\rm s,0}=0.01\,M_{\rm h}$.
Each plot is labeled with the value of the $x_{\rm c} (E)$.  We
identify the region of the parameters space where the satellite sinks
to the center of the main halo (M), evaporates in the background (D)
or survives (S).}
\label{fig:fate}
\end{figure*}
In Appendix B we give a simple expression for the disruption
time $t_{\rm dis}$ that can be used for a comparison with
the other timescale. 

\subsection  {Merging, distruption or survival}
We here investigate the dynamical
evolution of a {\it population} of satellites, in a given halo. An individual 
satellite is labeled by four
parameters; $x_{\rm c} (E)$
and $\cir$ identify the orbit, while initial
mass $M_{\rm s,0}$ and concentration $c_{\rm s}$ identify  the 
internal properties. Each combination of the four parameters leads to a
different final state for the satellite: rapid merging toward the
center of the main halo (M), disruption (D), or survival (S) (when a
residual mass $M_{\rm s}$ remains bound and maintains its identity,
orbiting in the main halo for a time longer than the Hubble time).

An important role is played by the concentration ratio as shown by the
life diagrams in Fig.~(\ref{fig:fate}).  These predict the final fate
of a satellite with $M_{\rm s,0}=0.01\,M_{\rm h}$, as function of the
$c_{\rm s}/\c$ and of the orbital parameters. The fractional area in
this parameter space leading to disruption, survival or decay is an
estimate of the relative importance of these processes in determining
the satellite fate. Disruption due to the tidal perturbation is the
fate of those satellites that initially move on close orbits despite
$c_{\rm s}/\c.$ Satellites moving along typical (plunging)
cosmological orbits survive over a Hubble time only if they had a
concentration higher than that of the main halo at the time of their
infall.

\begin{figure}
\centerline{\psfig{figure=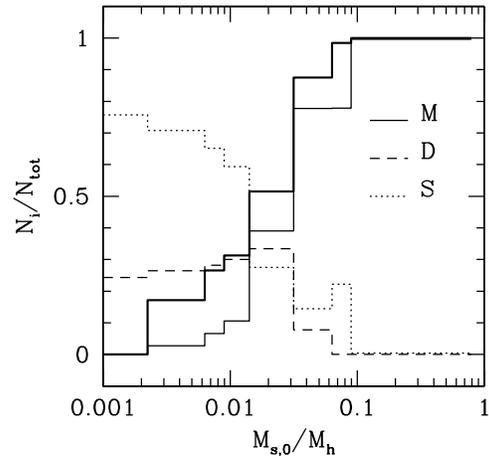,width=7.0cm}}
\caption[]{Probability distribution for the three final endpoints:
merging  (M),
disruption (D) or survival (S) as function of the initial satellite
mass. The thick solid line refers to the case of a rigid
satellite.}
\label{fig:parvol}
\end{figure}
%

In Fig.~(\ref{fig:parvol}) we have drawn the probability distribution
relative to the three final states: direct merging (by dynamical
friction), which dominates at large masses, survival and/or disruption
which is the most likely end for satellite with $M_{\rm
s,0}<0.01\,M_{\rm h}.$ In producing Fig.~(\ref{fig:parvol}) we have
generated evolutionary paths (ending after a time equal to the Hubble
time) for satellites starting from a flat distribution of orbital
parameters and concentrations.

\begin{figure}
\centerline{\psfig{figure=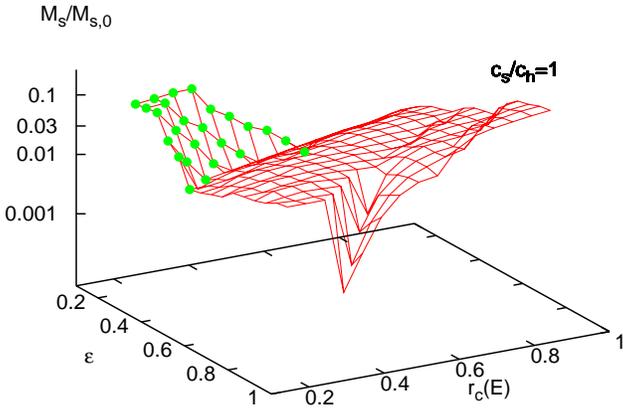,width=10.0cm}}
\caption[]{
The mass that remains 
as a function of the orbital parameters after more than a Hubble time,
for a satellite with $M_{\rm s,0} = 0.01 \, M_{\rm h}.$
Dots identify
the regions where the satellite is disrupted. The contour line on the
xy plane identify the loci where $M_{\rm s}/M_{\rm s,0}=0.1$.}
\label{fig:fatemas}
\end{figure}
\begin{figure}
\centerline{\psfig{figure=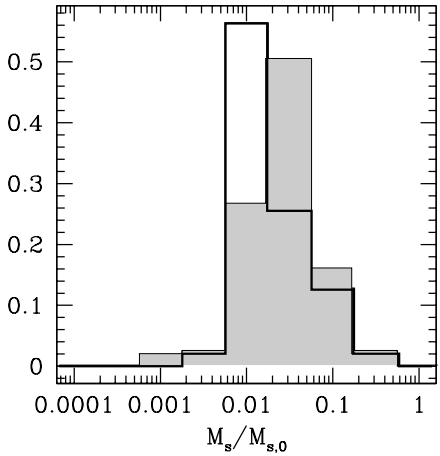,width=7.0cm}}
\caption[]{
The distribution of the final mass of a satellite 
of initial mass $M_{\rm s}= 0.01
M_{\rm h}$ after more than a Hubble time. Histograms 
are derived starting from 
a uniform distribution of orbital parameters, for two
values of the concentration ratio: $c_{\rm s}/\c=2$ (filled
grey area) and $c_{\rm s}/\c=1$.}
\label{fig:masdis}
\end{figure}

Our study suggests that those satellites that survive have  lost
memory of their initial state: dynamical friction 
perturbs the orbit and tidal stripping reduces the satellite
mass.  
In Fig.~(\ref{fig:fatemas}) we compute the mass of the 
satellites that remains after a Hubble time.
The figure refers to a high concentration case, but we
extend our analysis also to low concentrated satellites as shown in
Fig.~(\ref{fig:masdis}), where we compute the cumulative mass
distribution for all the initial orbital parameters.  On average, much less
then 10\%  of the initial mass remains bound. Of course, in
general circular orbits do not cause serious damages to
the satellite as shock heating is less
intense (an exception is represented by satellites on very 
tightly bound orbits).
In Fig.~(\ref{fig:fatemas}), dots show the final mass just prior
evaporation.  As expected, radial orbits can more easily dissolve a
satellite.

The strength of the orbital decay can be estimated measuring the
reduction of the apocenter distance.
\begin{figure}
\centerline{\psfig{figure=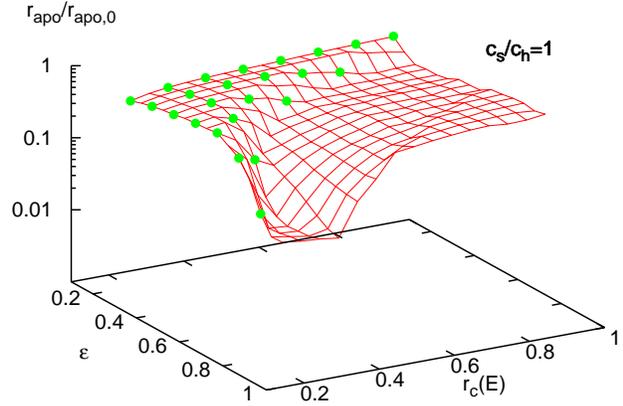,width=10.0cm}}
\centerline{\psfig{figure=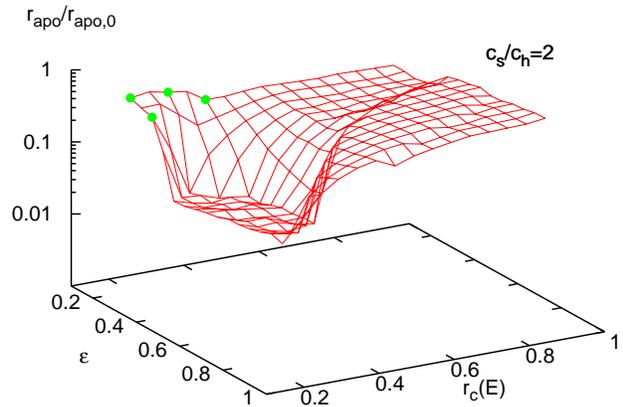,width=10.0cm}}
\caption[]{
The apocenters radius in units of its initial value for a satellite
$M_{\rm s,0}= 0.01 M_{\rm h}$ evaluated after more than a Hubble time
(we terminate at $15$ Gyr). The two plots refer to $c_{\rm s}/\c=2$
(bottom) and $c_{\rm s}/\c=1$ (top).  Dots identify satellites that
evaporate before 15 Gyr. The contour lines on the xy plane select the
regions where the relative reduction of $r_{\rm apo}$ is 0.1 (dotted
lines) and 0.5 (dashed lines).}
\label{fig:apofin}
\end{figure}
In Fig.~(\ref{fig:apofin}) we plot the distribution
of apocenter radii for a satellite
with $M_{\rm s,0}= 0.01 M_{\rm h}$ after 15~Gyr of orbital evolution. The
strength of the drag force reduces the apocenter distance by a factor of two
for cosmological orbits and it is not significantly affected by the
concentration. 

\subsection{Cosmological examples}
Now, we apply our analysis to some cosmological relevant examples.
We discuss the evolution of different 
satellites which orbit in a cluster-like and galaxy-like haloes.  The
cluster halo is a Coma-like cluster with  mass $M_{\rm h}=5\cdot
10^{15} M_\odot$ and concentration $\c=3.44$. The Milky Way-like galaxy 
halo has $M_{\rm h}=10^{12} M_\odot$ and $\c=10.44$.
For all cases, the initial orbital parameters are chosen as 
$\cir=0.6$ and $x_{\rm c}(E)=0.5$.

\vspace{0.5cm}
\noindent {\bf Group in Coma}

\noindent 
We consider a massive group-like satellite of mass $M_{\rm s,0}=3\cdot
10^{13} M_\odot$ and $c_{\rm s}=7.5$ which enters the Coma-like halo
at $z=0.5$. In a $\Lambda$CDM Cosmology it evolves for $\sim$4.8~Gyr
inside the halo until $z=0$.

As suggested by the high value of $c_{\rm s}/\c$ the satellite is not
disrupted. Since $M_{\rm s,0}= 0.006 M_{\rm h}$ the orbit is stable
and, with this choice of the initial orbital parameters, the satellite
evolves for $\sim$ 1.5 P$_{\rm orb}$. The final apocenter radius is
$r_{\rm apo}\simeq 0.85 r_{\rm apo,0}$ and its final mass is $M_{\rm
s}= 7.2 \cdot 10^{12} M_\odot$.

\vspace{0.5cm}
\noindent {\bf Milky Way in Coma}

\noindent
A Milky~Way-like satellite has mass $M_{\rm s,0}= 10^{12} M_\odot$ and
$c_{\rm s}=10.44$. If it enters the Coma-like halo at $z=0.5$ it
evolves for $\sim$ 1.5 P$_{\rm orb}$.  The orbit remains almost
unperturbed ($r_{\rm apo}\simeq 0.99 r_{\rm apo,0}$) as strength of
the drag force is extremely weak since $M_{\rm s,0}= 0.0002 M_{\rm
h}$. Due to the extremely high relative concentration, $c_{\rm
s}/\c\simeq 3$ the satellite does not evaporate and its final
mass ad $z=0$ is $M_{\rm s}= 2.5 \cdot 10^{11} M_\odot$

\vspace{0.5cm}
\noindent {\bf Large Magellanic Cloud in Milky Way}

\noindent
A Large Magellanic Cloud halo has $M_{\rm s,0}= 10^{11} M_\odot$ and
$c_{\rm s}=11.9$. As expected due to its relative high mass, the
satellite merges with the Milky Way in $\sim$ 4 Gyr. Before
merging, the satellite loses 97\% of its mass that is
dispersed in the Milky Way Halo.

\vspace{0.5cm}
\noindent {\bf Dwarf in Milky Way}

\noindent
We consider a Dwarf-like satellite of mass $M_{\rm s,0}= 5\cdot10^{9}
M_\odot$ and concentration $c_{\rm s}=13.6$.  If it enters the Milky
Way-like halo at $z=0.5$ it evolves on an almost unperturbed orbit for
$\sim$ 2 P$_{\rm orb}$ and its final mass is $M_{\rm s}= 2 \cdot
10^{8} M_\odot$.

\vspace{0.5cm}
\noindent {\bf Dwarf in Milky Way at high redshift}
\noindent
A Dwarf-like satellite enters a Milky Way-like halo at $z=2$, when the
Milky Way has mass $M_{\rm h}= 10^{11} M_\odot$ and concentration
$\c=6.15$. The satellite has $M_{\rm s,0}= 5 \cdot10^{9}M_\odot$
and {\bf $c_{\rm s}=6.8$}. The dwarf evolves for $\sim$11 Gyr. Due to its
low relative concentration it loses 99\% of its initial mass 
during the first orbital period; its orbit then  becomes stable
($r_{\rm apo}\simeq 0.36 r_{\rm apo,0}$). 
Note that here we are not accounting for the evolution of the main halo
which grows in mass during the remaining 11 Gyr before $z=0$,
but the influence of accreted mass on the central regions 
dynamics should be relatively small (Helmi et al 2002) 
Finally we notice that our Milky Way model does not account for the presence of a disk.
Penarrubia, Kroupa \& Boily (2002)
suggest that orbital evolution changes when the potential well has a  
flattened component, and  DF is more efficient 
for satellites with low-inclination orbits (respect to the disk):
the orbital decay is accelerated and the orbital plane decays over the 
disk plane. They also find instead that  DF enhances the survival time of satellites 
when they move on near-polar orbits.

\section{Summary and Conclusions}

We coupled together two successful models for dynamical friction and
tidal stripping and compared their predictions with high resolution
N--body simulations to address the evolution and ultimate fate of
satellite haloes within cosmic structures. Under the assumption that
haloes are well described by an NFW profile we are able to
predict if a satellite of given mass, orbital eccentricity  and infall
redshift, will merge, evaporate or survive under the simultaneous
action of dynamical friction and tidal mass loss.

We emphasize that we have obtained a complete predictive scheme
for the fate of a satellites whose masses 
at the time of infall into the main halo are known (below
we refer to typical cosmological orbits):

\begin{itemize}
\item
High mass satellites ($M_{\rm s,0} > 0.1 M_{\rm h}$) sink rapidly
toward the center of the main halo without significant mass dispersal:
The dynamical friction timescale for a rigid satellite
(equation [\ref{eq:fittau01}]) gives the correct timescale of merging.
\item
For satellites of mass $0.01 M_{\rm h} <M_{\rm s,0}<0.1 M_{\rm h}$
dynamical friction is still strong and drives the satellite toward the
center where tidal mass loss become severe.  Low concentration
satellites are disrupted, while high concentration satellites,
severely pruned by the tidal field, survive with masses $0.01 M_{\rm
s,0},$ and settle into inner orbits with a typical reduction of the
apocenter radii of a factor $\sim 0.1$ lower relative to the initial value.
The dynamical friction timescale for these satellites is longer than
for their rigid counterpart, and is given by equation~(\ref{eq:fittau001}).

\item
Light satellites with mass $M_{\rm s,0}< 0.01 M_{\rm h}$ are almost
unaffected by dynamical friction which is operating on a rather long
timescale. Mass loss by the tidal field, which is not severe on these
cosmological orbits, stabilizes further the orbit.

\item{Low concentration satellites below $0.1M_{\rm h}$ 
can be disrupted by tides before their orbital decay is
complete. Comparison of the dynamical friction timescale and the
disruption timescale as provided in this paper 
allows to find the actual lifetime of
satellite haloes.}

\end{itemize}

We predict that, because of the combined action of
stripping and dynamical friction, a primary halo at $z=0$ will host 
preferentially
satellites with mass $M_{\rm s}/M_{\rm h}\ll 0.01$ as the heavier
ones would have been accreted or/and dispersed in the background, leaving a
``depression'' in the mass function of substructure 
above 0.01 $M_{\rm s}/M_{\rm h},$(of course we are neglecting effects 
due to the evolution of the main halo itself). 
This feature should be more evident
in Milky Way-size haloes than in cluster haloes as in the former bound 
satellites had more time to evolve.

Since the destructive power of the tidal field (and in particular of tidal
shocks) depends sensitively on the degree of circularity of the
satellite orbit, a large galaxy halo like that of the Milky Way
($>10^{12} M_{\odot}$) should host satellites moving preferentially on
circular orbits as a consequence of the selective action of the tidal
field.  Also, because dynamical friction seems unable to render the
satellites' orbit circular (van den Bosch et al. 1999; CMG99),
 the low eccentricities should have been already
present as initial conditions.   This ``selection effect'' will
be extremely weak for smaller satellites (below $0.01-0.001 M_{\odot}$)
because their orbit barely decays and thus will have in general long
survival times (only low concentration satellites could disappear
quickly but they are not common in CDM models; see e.g. Eke, Navarro
\& Steinmetz 2001 and Bullock et al.2001). 
This mass regime corresponds to that of the 
dwarf spheroidal satellites of the Milky Way. On the other
end, the Magellanic Clouds, the dwarf elliptical satellites of M31 and
perhaps the dwarf spheroidal Fornax are all massive enough to fit in the
intermediate regime where destruction is
still possible; thus these galaxies could have survived because their 
host haloes had nearly circular orbits.
In the case of the Magellanic Clouds a nearly
circular orbit is indeed measured (Kroupa \& Bastian 1999).  There is,
however, at least one caveat to this interpretation, namely that both
the dwarf ellipticals of M31 and the Magellanic Clouds could
be dense enough to survive shocks on even very eccentric orbits
(Mayer et al. 2001b).  Only when all the orbits of the satellites will
be accurately determined we will know  whether eccentricity or
internal structure was more important in determining their survival.

The calculations described in this paper can become a useful tool when
coupled to cosmological simulations. The final goal is to find an
appropriate description of the dynamical evolution of substructure in
a halo.  Increasing computing power and code performances has recently
allowed to carry out extremely high resolution simulations that follow
the evolution of substructure in dark matter haloes (Ghigna et
al. 1998, 2000; Mayer et al. in preparation). These represent the new
ground where CDM models are being tested and their predictions
compared to observations.  However, these simulations remain costly
and usually only one system at a time can be simulated down to very
small scales. On the other end, resolving the mass function of
substructure in depth is important in light of the problem of the
overabundance of satellites (Moore et al. 1999; Klypin et
al. 1999). Such mass function can be viewed as the convolution of the
mass function of satellites at an earlier epoch with an evolutionary
filter function that depends on the dynamical mechanisms analyzed in
this work. Therefore, our results can allow to address the
substructure problem in a statistical way orders of magnitude faster
than with N-Body simulations; as an example one can explore a large
number of dynamical histories by randomly varying the orbital and
structural parameters in the range typical of cold dark matter
cosmogonies, and work is in progress (Taffoni et al. 2002 in
prep). Here we make a first attempt starting with uniform
distributions.  Clearly the time dependent potential of the growing
primary halo, whether it is a galaxy or a cluster, is an additional
ingredient that only simulations can incorporate and which could
affect the orbital dynamics of the satellites. However, the latter
limitation can be partially overcome by using the merger tree
extracted from a low resolution simulation, as done within some
semi-analytical models (Somerville \& Primack 1999; Kauffmann et
al. 1999; Cole et al. 2000) or using analytical merger trees providing
a good approximation to the latter (Taffoni et al. 2002; Monaco et
al. 2002).

A key result of our analysis, and one that is in agreement with the high
resolution cosmological simulations from which the initial orbits were
drawn, is that the inner, most bound part of small satellites as
concentrated as expected in CDM models (Eke, Navarro \& Steinmetz 2001;
Bullock et al. 2001) survive for timescales
comparable to or longer than the age of the Universe. This residual
has a size corresponding to a few percents of the initial virial
radius; this is comparable to the scale of the baryonic component in
galaxies, so we can argue that galaxies will mostly survive within the
main halo. This result is also confirmed by high-resolution SPH
simulations of the formation of Milky Way-like galaxy (Governato et
al. 2002).  Indeed, dissipation could make  the inner part of the haloes
even more robust against tides (Navarro \& Steinmetz 2000).  On the other
end, additional tidal shocks occurring during encounters between substructure
haloes, i.e. galaxy harassment (Moore et al. 1996, 1998), might have 
a counteracting effect and could actually increase mass loss. 
However, detailed simulations of this mechanism have shown that only very 
fragile, LSB-type galaxies would be severely damaged by harassment (Moore 
et al. 1999); halo profiles of these galaxies likely correspond to the
low concentration satellites studied in this paper (Van den Bosch
\& Swaters 2001) which we have shown are easily disrupted even by
the tides of the primary halo alone. Thus, adding harassment would only
accelerate the disruption of a few satellites while not affecting   
the survival of the majority of them which, in CDM models, 
have high concentrations. Hence the picture emerging from the life diagrams
of the satellites shown in this paper is robust.
Satellites close to disruption
at the present time, like Sagittarius in the Milky Way subgroup, must
have been much bigger in the past in order for dynamical friction to
drag them to an inner orbit where dissolution can easily take place; 
alternatively they could have entered the halo at fairly high redshift, 
which would place them naturally on an inner, tightly bound orbit (Mayer et
al. 2001b).  In clusters, dwarf galaxies cannibalized by giant CD
galaxies might also trace an early population.

Satellites infalling at redshift one or lower in the main haloes will
complete several orbits and  eventually undergo morphological changes 
by tidal stirring (Mayer et al. 2001a,b) and harassment (Moore et al.
1996, 1998). 
These will produce diffuse streams of stars while they
are orbiting (Helmi \& White 1999; Johnston, Sigurdsson \& Hernquist
1999), contributing to the build up of
an extended stellar halo population. Such
population should be present out to more than 200 kpc in the Milky Way
halo, as the plunging orbits of satellites seen in cosmological
simulations go this far out. On the contrary, a less extended stellar
halo should be expected if dynamical friction were more efficient in
dragging satellites to the center. The amount of stellar halo
substructure out to large distances could thus reveal the original
mass function of observed dwarf spheroidal galaxies in the Local
Group.  Components decoupled in their kinematics as well as in the
metallicity and age of their stars should be present, but tracking
such properties might be a daunting task observationally if enough
phase mixing occurs (Ibata et al. 2001a,b); however, while in the inner
halo fast orbital precession and heating by other clumps might blur
the streams, the phase space distribution of the outer halo material
should still carry the memory of the initial orbits of the satellites
(Mayer et al. 2002).

The prescriptions for the decay and disruption rates
obtained in this work provide a complete framework which can improve
the predictive power of semi-analytical models of 
galaxy formation. They enable to follow the complex evolution of 
substructures in hierarchical models in a straightforward manner.

\section{Acknowledgments}

The authors like to thank Tom Quinn and Joachim Stadel for
providing us  PKDGRAV. Thanks go to Marta Volonteri and Pierluigi 
 Monaco for useful discussions and to Valentina  D'odorico for the critical reading of the
manuscript.  Simulations have been carried out at the CINECA
Supercomputing Center (Bologna) and on a dual-processor ALPHA
workstation at the University of Washington. L. Mayer was supported by
the National Science Foundation (NSF Grant 9973209).


\appendix 
\section[]{Calculation of the tidal energy for a NFW profile}
At each pericenter passage the satellite crosses very rapidly the
central and more concentrated regions of the primary halo. The
duration of those encounters is fast compared with the dynamical time
of the object.  Such kind of interactions are called tidal shocks
(Spitzer 1987).  We will use the results derived by GHO to describe
the amount of heating due to tidal shocks on a satellite moving inside
an extended mass distribution.

During an orbital period $P_{\rm orb}$ the tidal force $\bld{f} _{\rm
s,tid}$ per unit mass produces a global variation on the velocity of
the internal fluid:
\beq
\label{eq:deltav}
\Delta {\bf v} = \int_{0}^{P_{\rm orb}} {\bf f}_{\rm tid} \;dt\;, 
\eeq 
where we have applied the impulse approximation in the hypothesis that
the time scale of interaction is short compared with the dynamical
time of the satellite ($t=0$ refers to the initial satellite position
at apocenter distance).  

In a spherically symmetric system of mass $M_{\rm h}$, the tidal force
per unit mass exerted by the background on a dark matter particle of
the satellite is:
\beq
\label{eq:tforce}
{\bf f}_{\rm tid} = \frac{GM_{\rm h}}{R_{\rm h}^3}
[(3\mu-\mu')(\hat {\bf r}\cdot {\bf R_{\rm s}})\hat 
{\bf r} -\mu {\bf R_{\rm s}}]\;,
\eeq
where $\hat {\bf r}={\bf r}/R_{\rm h}$ is the direction to the center
of mass of the satellite (CMS), $\bld R_{\rm s}$ is the position of the
particle respect to CMS. Note that $R_{\rm h}$ is the virial radius of
the main system. Here:
\beq
\mu(r)=\frac{M(r)}{M_{\rm h}}
\eeq
is the adimensional mass profile, and: 
\beq
\mu'(r)=\frac{d\mu(r)}{d\ln r}\;.
\eeq
For a NFW profile $\mu$ and $\mu'$ are functions of the normalized
radius $x=r/R_{\rm h}$ and of the concentration $\c$ of the
primary halo:
\beq
\mu(x,\c)={\ln(1+\c x)-\c x/(1+\c x) \over  \ln(1+\c)-\c/(1+\c)}\;,
\eeq
and
\beq
\mu'(x,\c)={1 \over
\ln(1+\c)-\c/(1+\c)} \left( {\c x \over 1+\c x} \right)^2 \;.
\eeq 

In the case of stable orbits the angular momentum $J$ is conserved and
we can use the identity 
\beq
\label{eq:a:var}
dt=(r^2/J)d\theta
\eeq 
to re-write
equation(\ref{eq:deltav}) into components (GHO):
\beq
\label{eq:shock}
\Delta {\bf v} =  {G M_{\rm h} \over r J} 
\{(B_1-B_3)x,(B_2-B_3)y,-B_3 z\}\;,
\eeq
where
\begin{eqnarray}
\label{shocktot}
B_1(\c) & = & \int_{-\theta_m}^{-\theta_m} \, F_1(x,\c) \,\cos^2{\theta} \, d\theta \\
B_2(\c) & = & \int_{-\theta_m}^{-\theta_m} \, F_1(x,\c) \, \sin^2{\theta} \, d\theta \\
B_3(\c) & = & \int_{-\theta_m}^{-\theta_m} \, {\mu(x,\c) \over x} \, d\theta,
\end{eqnarray}
where $\theta_m$ is the maximum value of the position angle, and:
\bea
\nonumber
F_1(x,\c)= 3
\eea
\beq
\times{[\ln(1+\c x)-\c x/(1+\c x)]-[\c x/(1+\c x)]^2 \over x
[\ln(1+\c)-\c/(1+\c)]}\;.
\eeq
This velocity changes cause a reduction of the binding energy of the
system:
\beq
\label{eq:deltaE}
\langle \Delta E \rangle= \left \langle \frac 1 2 |\Delta 
{\bf v}|^2  \right \rangle\;.
\eeq

Averaging over an ensemble of dark matter particles in a spherically
symmetric satellite we have that $\langle x^2
\rangle= \langle y^2 \rangle=\langle z^2 \rangle=R_{\rm s}^2 \:/ \:2$, 
and using equation(\ref{eq:shock}), the tidal energy gained by the
satellite becomes:
\bea
\nonumber
\langle \Delta E \rangle&=& \left({G M_{\rm h} \over
J \, R_{\rm h} } \right)^2  \\
&\times& \left[ { (B_1-B_3)^2+(B_2-B_3)^2+B_3^2
\over 6} \right] \; R_{\rm s}^2 \;.
\eea
In the previous expression, the contribution due to the
halo and the orbital parameters ($\cir$ and $x_c^2[E]$) is confined in
the function:
\bea
\nonumber
\Xi\left[\c,x_{\rm c}(E),\cir\right] &=&
\left({G M_{\rm h} \over R_{\rm h}^2 V_{\rm h}} \right)^2 
{1 \over  x_{\rm c}^2(E) \, \cir^2} \\
&\times& \left[ { (B_1-B_3)^2+(B_2-B_3)^2+B_3^2 \over  6 } \right] \;,
\eea
here $V_{\rm h}$ is the circular velocity of the main halo at virial
radius. 

It is then useful to write the shock energy as:
\beq
\langle \Delta E \rangle =
\Xi\left[\c,x_{\rm c}(E),\cir \right] 
\; R_{\rm s}^2
\;.
\eeq
\begin{figure}
\centerline{\psfig{figure=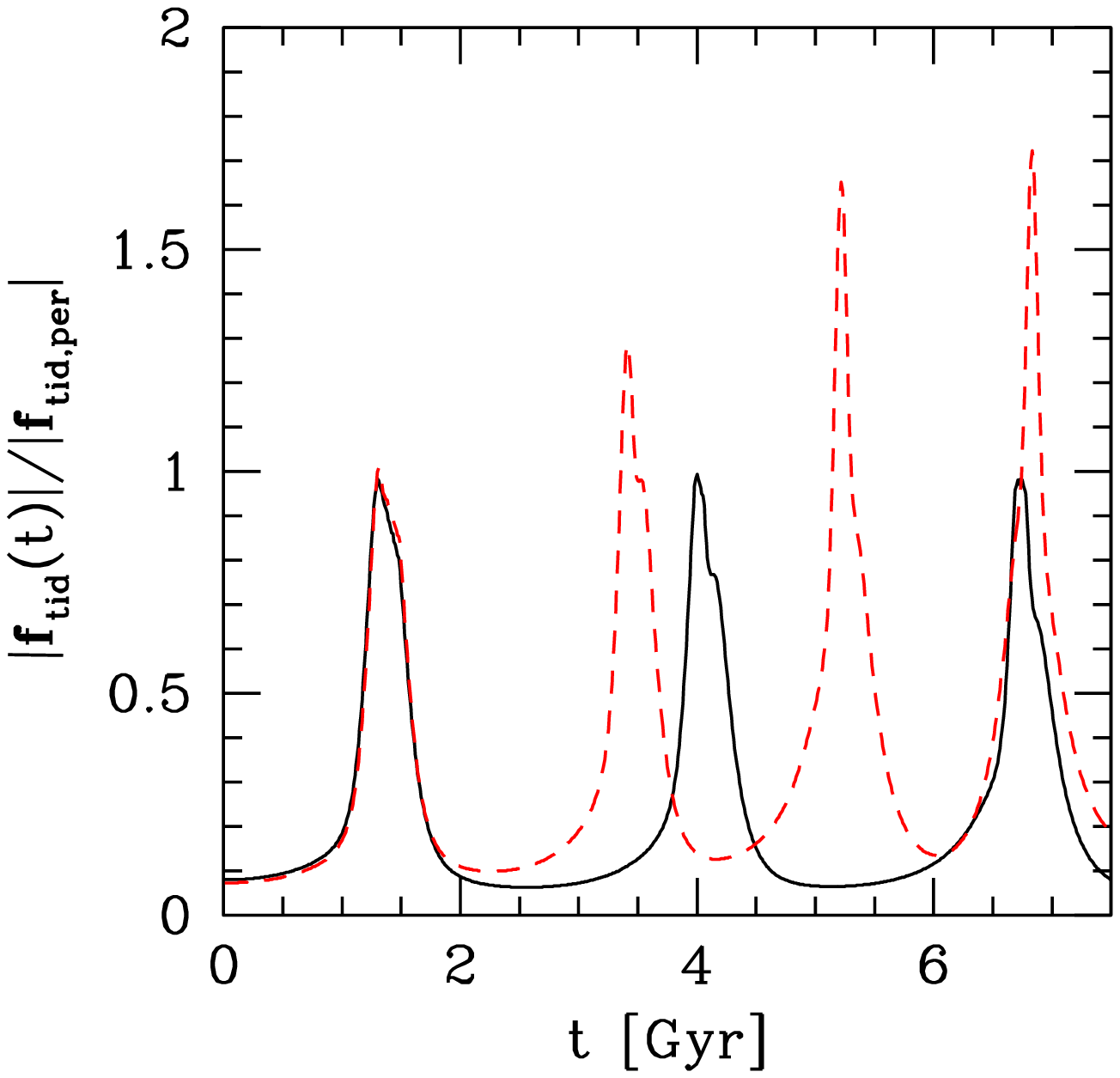,width=8.0cm}}
\caption[]{
The intensity of the tidal force $|{\bf f}_{\rm tid}(t)|$ normalised to
its value at the first periastron $|{\bf f}_{\rm tid,per}|$.  We plot
the module of the tidal force (equation~[\ref{eq:tforce}]) as a function
of time for a satellite of mass $M_{\rm s,0}=0.01 M_{\rm h}$ and
$\cir=0.7$ $x_{\rm c}(E)=0.5$.  The solid line refers to a stable
orbit, the dashed line to an unstable one. When the drag force
is active, the intensity  of the tidal force, and consequently
of the shock energy, grows with time.}
\label{fig:ftid}
\end{figure}
When the frictional drag force is active, it is not possible to change
the integration variable according to equation~(\ref{eq:a:var}). The
energy change becomes:
\bea
\nonumber
\langle \Delta E \rangle &=&  
\left({G M_{\rm h} \over R_{\rm h}^3 } \right)^2  \\
\label{eq:shockNON}
&\times& \left[ { (A_1-A_3)^2+(A_2-A_3)^2+A_3^2 \over 6} \right] 
\; R_{\rm s}^2 \;.
\eea   
Here
\bea
\label{shocktot1}
A_1(\cir,x_{\rm c}[E])  &=&  \int_{0}^{P_{\rm orb}} \, F_2(x,\c) \,\cos^2{\theta} \, dt \\
A_2(\cir,x_{\rm c}[E])  &=&  \int_{0}^{P_{\rm orb}} \, F_2(x,\c) \, \sin^2{\theta} \, dt \\
A_3(\cir,x_{\rm c}[E]) &=&  \int_{0}^{P_{\rm orb}} \, {\mu(x,\c) \over x^3} \, dt,
\eea
with
\beq
F_2(x,\c)= {F_1(x,\c) \over x^2}
\eeq
Once again we separate the contribution due to the environment:
\bea
\nonumber
{\cal F}[\c , x_{\rm c}(E), \cir  ] &=&
\left({G M_{\rm h} \over R_{\rm h}^3 } \right)^2 \\
&\times&
\left[ { (A_1-A_3)^2+(A_2-A_3)^2+A_3^2 \over 6} \right] \;.
\eea
For an unstable orbit 
\beq
\langle \Delta E \rangle =
{\cal F}\left[\c,x_{\rm c}(E),\cir\right] 
\; R_{\rm s}^2
\;.
\eeq 
The shock energy in this case must be evaluated along the perturbed
orbit. As the drag force drives the satellite in the internal region
of the halo the $\langle \Delta E \rangle$ increases
(Fig.~[\ref{fig:ftid}]).
\section[]{An approximate estimate of the disruption timescale}

Because the lifetime of light satellites is mostly set by tidal
disruption we estimate here the disruption time. If the main halo
density profile is isothermal (ISO) then GHO showed that the shock
energy change is:
\beq
\label{eq:nfw.iso}
\langle \Delta E \rangle_{\rm ISO}= \left(V_{\rm h} \over
R_{\rm h} \right)^2 { 2 \sin^2\theta_{\rm m} + 4 \theta^2_{\rm m} 
 \over 6(\epsilon \; x_{\rm c} )^2} A(x_\tau) \cdot R_{\rm s}^2 \;,
\eeq
where $\theta_{\rm m}$ is the maximum value of the position angle
which varies from $\pi/2$ to $\pi$.  Using the orbit equation
(equation~[\ref{eq:orbiteq}]) we can evaluate $\theta_{\rm m}$:
\beq
\theta_{\rm m}=2 \, \epsilon \, x_{\rm c}  
\int_{r_{\rm per}}^{r_{\rm apo}} {dx \over
x^2\sqrt{\ln(r_c/x)^2-(r_c/x)^2\epsilon^2-1}}
\eeq
and the orbital period
\beq
P_{\rm orb}=2 {R_{\rm h}\over V_{\rm h}} \,  
\int_{r_{\rm per}}^{r_{\rm apo}} {dx \over
\sqrt{\ln(r_c/x)^2-(r_c/x)^2\epsilon^2-1}}\;.
\eeq
	
The shock in the ISO profile equals the shock of an NFW case when
$\c=30$. We have than:
\beq
\langle \Delta E \rangle_{\rm NFW} \sim \langle \Delta E \rangle_{\rm ISO}
\times (0.029\;\c+0.13)\;.
\eeq

At each pericenter passage the satellite is shock heated and its radius
$R_{\rm s}$ is reduced of a factor $\Delta R$. As an  approximation
$\Delta R \sim R_{\rm s,0}/N$ where $N$ is the number of pericenter passages 
necessary to destroy the satellite.
Then, we have an implicit equation for $N$
\beq
\label{eq:appB:1}
N+\frac 1 N \sum_{i=1}^{N-1} i^2={E_0 \over \langle \Delta E
\rangle_{\rm NFW,0}}\;,
\eeq
where $E_0=0.5 G M_{\rm s}/R_{\rm s,0}$, and $\langle \Delta E
\rangle_{\rm NFW,0}$ is evaluated at the initial half mass radius. 
Since 
\beq
\sum_{i=1}^{N-1} i^2={N\,(N-1)\,(2N-1) \over 6}
\eeq
for large N equation~(\ref{eq:appB:1})  would give:
\beq
N \approx \sqrt{4 8 {E_0 \over  \langle \Delta E
\rangle_{\rm NFW,0}}}\;.
\eeq
The disruption time can then be written as:
\beq
\label{eq:shoprop}
t_{\rm dis}\sim P_{\rm orb}\cdot N\;.
\eeq
This formula provides a simple estimate of the disruption time
valid on cosmological relevant orbits with a precision of  
25 per cent. 

\label{lastpage}
\end{document}